\documentclass[
twocolumn,
amsmath,
amssymb,
aps,
prb,
eqsecnum
superscriptaddress,
floatfix,
english
]{revtex4-2}

\usepackage{amsmath}
\usepackage{amssymb}
\usepackage{graphicx} 
\usepackage{dcolumn}  
\usepackage{bm}       
\usepackage{hyperref} 
\usepackage{color}

\definecolor{amaranth}{rgb}{0.9, 0.17, 0.31}

\begin{document}
	
	\title{Transport, refraction and interface arcs in junctions of Weyl semimetals}
	
	\author{Francesco Buccheri} \email{buccheri@hhu.de}
	\affiliation{Institut f\"ur Theoretische Physik,
		Heinrich-Heine-Universit\"at, D-40225  D\"usseldorf, Germany}
	\author{Reinhold Egger} \email{egger@hhu.de}
	\affiliation{Institut f\"ur Theoretische Physik,
		Heinrich-Heine-Universit\"at, D-40225  D\"usseldorf, Germany}
	\author{Alessandro De Martino}
	\email{Alessandro.De-Martino.1@city.ac.uk}
	\affiliation{Department of Mathematics, City, University of London, Northampton Square, EC1V OHB London, United Kingdom}
	\affiliation{Dipartimento di Fisica "E. Pancini", Università di Napoli “Federico II", Complesso 
		di Monte S. Angelo, Via Cintia, I-80126 Napoli, Italy}
	
	\begin{abstract}
		We study the low-energy single-electron transport across a junction of two magnetic Weyl semimetals, 
		in which the anisotropy axes are tilted one respect to the other. Using a two-band model with a potential step, we compute the transmission factor for normal and Klein tunneling and the refraction properties of the interface as a function of the tilt angle. We show that the interface acts as a beam splitter, separating electrons with different chiralities.
		We also characterize interface states, only appearing for finite tilt angle, which connect the projection of the Fermi surfaces on the two sides of the junction, and we discuss transport effects due to their interplay with Fermi arcs.
	\end{abstract}
	
	\maketitle
	
	\section{\label{intro}Introduction}
	Weyl semimetals are three-dimensional materials, in which the valence and conduction bands are well-separated everywhere in the Brillouin zone, except at a finite number of isolated points, dubbed Weyl nodes. Here, two non-degenerate, approximately linear, bands cross, which make the sample behave as a semimetal when the chemical potential is approximately at the band crossing and originate the characteristic topological magneto-electric response \cite{Armitage2018,Vazifeh2013,Zhou2013,Volovik2009,Spivak2016}. 
	In many ways, this class of materials can be thought of as the three-dimensional analogue of graphene, but
	the extra dimension provides robustness against time-reversal breaking perturbations and a very large magnetic field is necessary in order to gap out the electronic spectrum \cite{Nielsen1981,Kim2017}. In addition, the topological protection of the band crossings provides robustness against moderate disorder \cite{Buchhold2018}. 
	
	One way for non-degenerate band crossings to appear in the spectrum is to break the time-reversal symmetry: Weyl semimetals of this family are dubbed magnetic and exhibit partial or full magnetization of the carriers at the Fermi level. 
	They provide an excellent playground for theorists and experimentalists due to the richness of exotic features, and
	are currently object of widespread attention because their bulk Berry curvature potentially allows extensive manipulation of electronic currents \cite{Yan2017,Zhou2019,daSilvaNeto2019}. They host surface states in the form of Fermi arcs, which are connected to nontrivial Hall response, magnetoconductance and thermal transport phenomena \cite{Burkov2011,Baireuther2016,Burrello2019}.
	Compounds in the pyrochlore iridates family and the ferromagnet $\mbox{HgCr}_2\mbox{Se}_4$ were the first candidate Weyl semimetals \cite{Wan2011,Xu2011}, followed by various layered materials \cite{Wang2016}
	, such as $\mbox{Co}_3\mbox{Sn}_2\mbox{S}_{2}$ \cite{Morali2019,Liu2019,Liu2021,Rossi2021} 
	and the antiferromagnets $\mbox{Mn}_3\mbox{X}$ (X=Sn, Ge) \cite{Yang2017}, in which the Weyl nodes were identified.
	Magnetic Heusler alloys also provide excellent candidates for time-reversal-broken Weyl semimetals \cite{Wang2016,Wollmann2017,Yan2017}, with various experimental confirmations, including \mbox{$\mbox{Ti}_{2}$MnAl} \cite{Feng2015,Shi2018,Esin2020}, $\mbox{Co}_2\mbox{TiX}$ (X=Si, Ge, Sn), GdSI \cite{Nie2017} and  $\mbox{Co}_{2}\mbox{MnGa}$ \cite{Belopolski2019}. See also \cite{Bernevig2022} for a recent review.
	
	The mounting number of experiments and the high degree of manipulation available on the samples prompts theoretical efforts to investigate systems with more complicated geometries. In this work, we address electronic transport at an interface between two magnetic Weyl semimetals which are tilted one with respect to the other. The relativistic spectrum is at the origin of intriguing transport properties, such as Klein tunneling \cite{Li2016,Yesilyurt2016}. 
	The helicity of the quasiparticles near the Weyl nodes, together with the spin texture of the Fermi arcs, also originates non-trivial physics at the interface with a normal metal \cite{Zhu2020} or with a superconductor\cite{Madsen2021}.
	A common instance of an interface may be created by imperfections in the sample, in the form of irregular surfaces or adjacent extended regions with misaligned lattice structures. Moreover, when a crystalline sample is abruptly cooled down, small cracks in the material can be generated, so that the lattice vectors are not perfectly aligned anymore on the two sides of the defect. Finally, a sharp domain wall between two regions with different magnetization \cite{Lee2022,Shen2022} can be described within our formalism \cite{Kobayashi2018}. In these situations, translation invariance in the direction perpendicular to the interface is broken.
	As a result of the displacement of the crystal axes, the dispersion of the electron changes across the interface, resulting in the refraction of an incident electron beam \cite{Yang2019}.
	A junction of two different non-centrosymmetric Weyl semimetals of the same family has been considered in \cite{Hills2017}: following the idea of Veselago lensing in graphene \cite{Veselago1968,Cheianov2006,Cheianov2007} the authors proposed application to scanning tunneling microscopy and to fine control of electron transport using a square potential barrier.
	\begin{figure}
		\centering
		\includegraphics[width=\columnwidth]{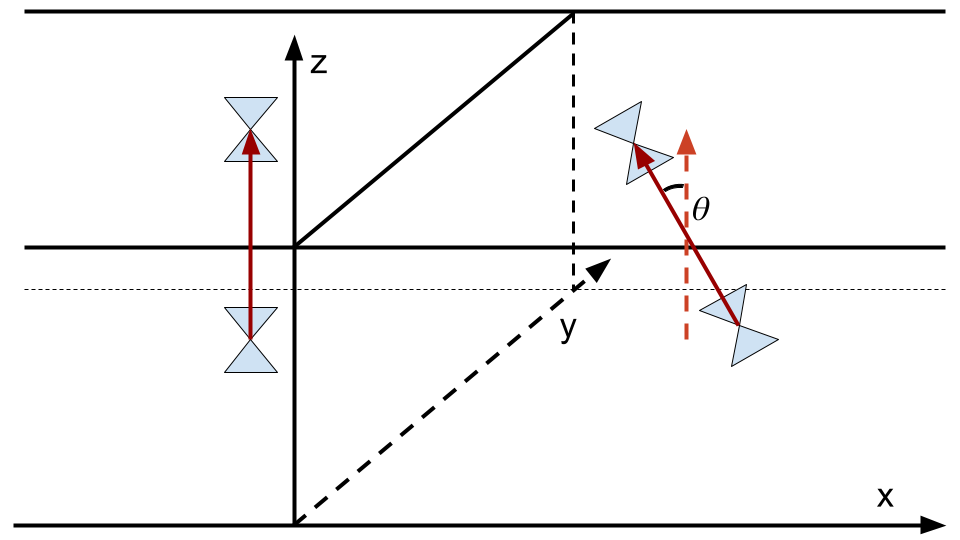}
		\caption{Two identical Weyl semimetals, with anisotropy axes rotated in the $yz$ plane by a tilt angle $\theta$, share a common interface at $x=0$. The system is infinite in the $y$ and $z$ directions.}
		\label{fig:model}
	\end{figure}
	
	We focus in this paper on a single interface and characterize its transmission properties as a function of the tilt angle, see Fig. \ref{fig:model}, and their impact on electric transport and thermoelectric properties of the sample. We also consider the possibility of different doping levels on the two sides of the sample, hence, the possibility of Klein tunneling, and compute how the interface affects the direction of a transmitted electron or hole. In addition, we  address the question of electronic states localized at the interface. For finite tilt, the projections of the Weyl nodes from the two sides do not coincide, which suggests the necessity of interface states joining the disconnected parts of the Fermi surface. A transfer matrix formalism has previously been applied in \cite{Dwivedi2018} to investigate how Fermi arcs on different sides of an interface can connect. The fate of surface states when two Weyl semimetals are tunnel-coupled has been further studied in \cite{Ishida2018}, see also \cite{Kobayashi2018,Abdulla2021} for the effect of a twist. In our work, we consider the transparent limit and find interface arcs, supporting chiral transport along the interface. We emphasize the universal origin of these interface states, whose main features can be derived using a simple low-energy model of a magnetic Weyl semimetal with a minimal pair of nodes.
	
	The article is structured as follows. In section \ref{sec:Model} we introduce the model underlying our study. In section \ref{sec:Transmission} we analyze electron tunneling and Klein tunneling across a transparent junction, and we determine the single-electron transmission amplitude as a function of the incoming momentum. 
	We then compute the low-temperature limits of the conductance and the thermopower per unit surface, 
	showing their relation with the shape and position of the projected Fermi surfaces. 
	In section \ref{sec:Refraction} we determine the refraction properties of the junction, 
	showing that the interface acts as a beam splitter, where electrons with opposite chiralities 
	are transmitted with different angles.
	In section \ref{sec:Interfacestates} we show that states localized at the interface, 
	in the shape of arcs in the Brillouin zone, are generically expected in a minimal continuum 
	model where the interface is transparent and the Fermi surfaces on the two sides are not 
	overlapping. 
	In section \ref{sec:Slabs} we introduce the contribution of Fermi arcs in the scattering 
	problem and argue that the interface states play an essential role in transport at large rotation angles. We offer conclusions and an outlook in section \ref{sec:Conclusions}. 
	Technical details about rotations, interface states and the diagonalization of the slab problem are provided in several Appendices.

	\section{Model}\label{sec:Model}
	
	A simple model for a Weyl semimetal with broken time reversal symmetry 
	is described by the Hamiltonian \mbox{($\hbar=1$)} \cite{Vazifeh2013,Okugawa2014,Gorbar2016,Burrello2019}
	\begin{equation}\label{HW0}
	H_0 = v k_x \sigma^x +  v k_y \sigma^y  +m\left(k_z\right) \sigma^z,
	\end{equation}
	where $v$ is the Fermi velocity and the Pauli $\sigma$ matrices act on a band (pseudospin) degree of freedom. 
	The "mass"
	\begin{equation}\label{m}
	m\left(k_z\right) = \frac{v}{2k_W} \left(k_z^2 -k_W^2\right)
	\end{equation}
	changes sign at $k_z=\pm k_W$ and singles out the $z$~axis as the anisotropy axis.
	The Hamiltonian \eqref{HW0} can be seen as a small-momentum expansion of a 
	widely used minimal two-band Hamiltonian of a magnetic Weyl semimetal \cite{Bovenzi2018,Armitage2018}.
	Then, two Weyl nodes with linear dispersion are present in the Brillouin zone at the momenta 
	$(0,0,\eta k_W)$, $\eta=\pm 1$.
	To each of these points it is possible to associate a  "chirality" $\eta=\pm 1$, 
	which coincides with 
	the quantized flux of the Berry curvature through a closed surface surrounding the node (divided by  \mbox{$2\pi$}).
	
	We will model an extended region with a lattice tilted with respect to the 
	adjacent one by starting from the Hamiltonian \eqref{HW0} and applying a rotation 
	of an angle $\theta$ around the $x$~axis, which rotates the internal "pseudospin" degree of freedom as well as 
	the anisotropy axis. 
	Denoting $\mathbf{k}=(k_y,k_z)^T$ the component of the momentum in the
	$yz$~plane and $\mathbf{k}_\theta=(k_{\theta,y},k_{\theta,z})^T=R_\theta \mathbf{k}$ its rotated counterpart,
	the transformed Hamiltonian is written as (see App.~\ref{sec:rotations})
	\begin{equation}\label{HWtheta}
	H_\theta = vk_x \sigma^x +\mathbf{b}_\theta \cdot \bm{\sigma},
	\end{equation}
	where 
	\begin{equation}\label{b}
	\mathbf{b}_\theta = \left(\begin{array}{c}
	b_{\theta,y}\\
	b_{\theta,z} 
	\end{array}\right)
	= R_{-\theta}
	\left(\begin{array}{c}
	v k_{\theta,y} \\
	m\left(k_{\theta,z} \right)
	\end{array}\right).
	\end{equation}
	The matrix $R_\theta$ represents the two-dimensional rotation in the $yz$ plane. 
	Note that if the mass function is chosen to be of the form $m(k_z)=v k_z$, the Hamiltonian and the spectrum are invariant under rotations. Conversely, with our choice of $m$ in \eqref{m}, the spectrum of \eqref{HWtheta} is composed by an electron ($\nu=+1$) and a hole ($\nu=-1$) branch, with eigenvalues
	\begin{equation}\label{Etheta}
	E = E_{\theta,\nu}\left(k_x,\mathbf{k}\right) = \nu\sqrt{v^2k_x^2 + \mathbf{b}_\theta^2}\;,
	\end{equation}
	corresponding to the eigenvectors given in Appendix \ref{sec:rotations}.
	It is readily checked at this point that the Weyl nodes are moved to the positions
	\begin{equation}\label{rotkW}
	{\bf k}_{W,\theta}^{(\eta)} =  \eta  k_W
	\begin{pmatrix} 
	-\sin\theta \\
	\cos\theta  
	\end{pmatrix}.
	\end{equation}
	The chirality associated with the node, instead, does not change with $\theta$.
	In the vicinity of the nodes the vector $\mathbf{b}_\theta$ assumes the simple linearized form 
	\mbox{$\mathbf{b}_\theta \approx v\left(\mathbf{k}-{\bf k}_{W,\theta}^{(\eta)}\right)$}.
	
	In this work, we consider an interface in the $x$ direction between two identical samples, 
	whose anisotropy axes are tilted by an angle $\theta$, see Fig. \ref{fig:model}.
	We also introduce a potential step in the form \mbox{$V(x)=\mbox{sgn}(x)V_0$}, with $V_0>0$, thus realizing a model \mbox{$np$-junction}. This configuration can be achieved by different doping levels or, for a sufficiently thin sample in the $z$ direction, via suitable gating.
	In real samples, a mismatch in the lattice orientation results, in general, in a larger inter-layer distance and in an only partially transparent interface. This can be modeled by a small region $-\ell<x<\ell$ around the origin in which a potential barrier of height $V_1\gg V_0, E$ is inserted. While this affects the transparency of the interface, no qualitative changes to the transmission coefficients are introduced in the thin barrier limit and, for the sake of simplicity, we omit this effect altogether. 
	We therefore write the Hamiltonian
	\begin{eqnarray}\label{Hint}
	H_\mathbf{k}=
	\begin{cases}
	v \hat{k}_{x} \sigma^{x} + v k_y \sigma^y+m\left(k_z\right)\sigma^z-V_0\;, & x<0\\
	v \hat{k}_{x} \sigma^{x} + {b}_{\theta,y}\sigma^y + b_{\theta,z} \sigma^z+V_0\;, & x>0
	\end{cases}
	\end{eqnarray}
	where $\hat{k}_x=-i\partial_x$.
	The tilt results in a mismatch in the position of the Weyl nodes on the two sides of the interface,
	with the displacement between the tilted and untilted Weyl nodes given by
	\begin{equation}\label{DeltakW}
	\Delta \mathbf{k}_W = 
	{\mathbf k}_{W,0}^{(\eta)}  -   {\mathbf k}_{W,\theta}^{( \eta)}  
	=2\eta k_W \sin \frac{\theta}{2}  
	\begin{pmatrix}
	\cos \frac{\theta}{2} \\ \sin \frac{\theta}{2} 
	\end{pmatrix} 
	\end{equation}
	for the nodes with the same chirality. For nodes of opposite chirality,
	the displacement is given by Eq.~\eqref{DeltakW} with $\theta$ replaced by $\theta+\pi$.
	The separation between tilted and untilted nodes plays an important role in 
	the scattering and refraction properties of the interface, which we will address in the following sections.
	
	The original Hamiltonian~\eqref{HW0} has inversion symmetry, rotational symmetry in the $xy$ plane, and
	particle-hole symmetry. For the interface Hamiltonian in Eq.~\eqref{Hint}, 
	all these symmetries are broken by the tilt between the left and right subsystems 
	and by the potential step.
	We note that the reflection in the plane 
	spanned by $\hat x$ and $R_{-\frac{\theta}{2}}\hat z$ exchanges the Weyl nodes on the two sides of the system. We show in App.~\ref{sec:symmetry} that this reflection is a 
	symmetry of the dispersion relation of the interface arcs if $V_0=0$.
	
	\section{Single-electron transmission and transport across an interface}\label{sec:Transmission}
	
	In this section, we study the scattering problem on the interface, in the presence 
	of a tilt $\theta$ and a potential step $2V_0$.
	In the regime $E>V_0$, an electron incoming from the left can be either reflected or
	transmitted as an electron through the potential step and one writes a scattering state in the form
	\begin{eqnarray}\label{psiscatteee}
	\psi_{E,\mathbf{k}}(x)	=	\begin{cases}
	u_{0;k_x,\mathbf{k}}e^{ik_{x}x}+r \,u_{0;-k_x,\mathbf{k}}e^{-ik_{x}x} & x<0 \\
	C\,u_{\theta;\bar{k}_{x},\mathbf{k}}e^{i \bar{k}_{x}x} & x>0
	\end{cases}
	\end{eqnarray}
	with the bulk eigenstates 
	$u_{\theta;k_{x},\mathbf{k}}$ given in \eqref{bulkeigenvec} and complex-valued coefficients $r$ and $C$.
	The momentum in the $x$ direction is not conserved, but depends instead on the energy via the relations
	\begin{eqnarray}\label{psiscatt}
	k_{x}&=&\frac{1}{v}\sqrt{\varepsilon_+^{2}-v^2k_{y}^{2}-m^{2}\left({k}_{z}\right)},\label{kx}
	\\
	\bar{k}_{x}&=&\frac{1}{v}\sqrt{\varepsilon_-^{2}-v^2{k}_{\theta,y}^{2}-m^{2}\left({k}_{\theta,z}\right)},
	\label{kthetax}
	\end{eqnarray}
	where $\varepsilon_\pm=E\pm V_0$. 
	In the regime $-V_0 \le E < V_0$, we consider an electron incoming from $-\infty$, 
	which can be either reflected as an electron or transmitted as a hole (Klein tunneling). 
	In this situation, the transmitted hole traveling in the positive $x$ direction is 
	described by a scattering state analogous to \eqref{psiscatteee}, with the wave function 
	in the region $x>0$ replaced by $u_{\theta;-\bar k_{x},\mathbf{k}}e^{-i \bar k_{x}x}$. 
	In the same way, one can describe the transmission and reflection of holes by changing appropriately
	the sign of $k_{x}$ and $\bar k_{x}$.
	\begin{figure}[h!]
		\centering
		\includegraphics[width=0.8\columnwidth]{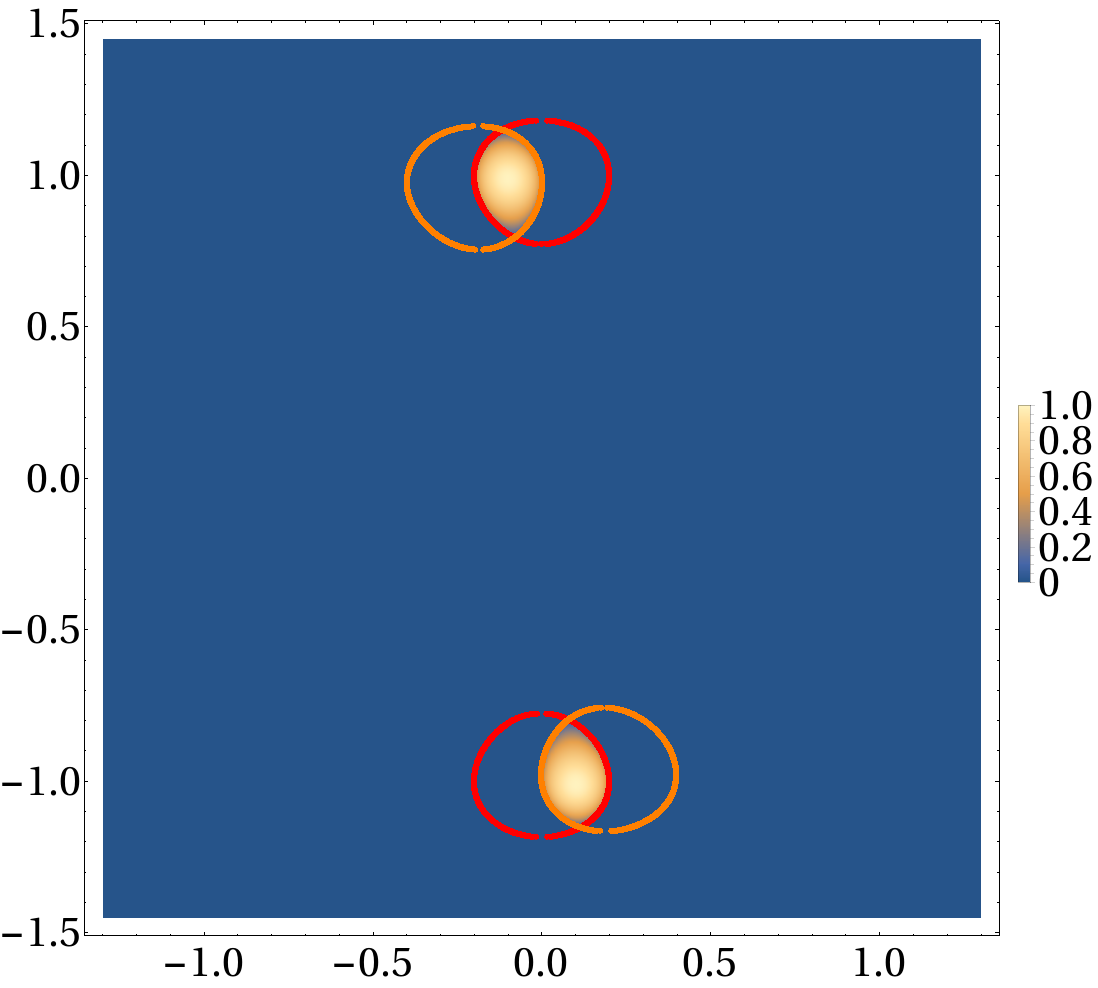}
		\caption{Intensity plot of the transmission factor \eqref{T} in the $k_yk_z$ plane (momenta in units of $k_W$) for a zero-energy electron with a potential step $V_0=0.2\,vk_W$ and tilt angle $\theta=0.2$. The red and orange circles delimit the Fermi surfaces on the two sides of the interface.}
		\label{fig:T1}
	\end{figure}
	
	Imposing the continuity of the wave function at the interface, the transmission probability ${\cal T}_{\theta}=1-\left|r\right|^2$
	follows as
	\begin{eqnarray}\label{T}
	{\cal T}_{\theta}\left(E,\mathbf{k}\right)&=&
	\frac{4k_{x} \bar k_{x}\mathcal{A}_{\theta}}
	{\left(\mathcal{A}_{\theta} k_{x}+ \bar k_{x} \right)^2 +\left(\mathcal{A}_{\theta}k_{y}-\nu_1\nu_2 b^{y}_\theta/v \right)^2},
	\end{eqnarray}
	where
	\begin{equation}
	\mathcal{A}_{\theta}=
	\frac{\left|\varepsilon_{-}\right|-\nu_{2} b_{\theta}^{z}}
	{\left|\varepsilon_{+}\right|-\nu_{1} m\left(k_z\right)}.
	\end{equation} 
	Here $\nu_1=\pm$ labels the particle/hole branch on the left of the interface and $\nu_2=\pm$ on the right. One can readily check that the transmission probability from right to left has the same expression.
	
	When the chemical potential is close to the band crossing, for $V_0\approx 0$, the reduced Fermi surface implies that very few states are available for transport. In particular, for $\mu \ll vk_W$, one can approximate it with a sphere of radius $\mu$ centered around each node. Hence, the Fermi surfaces on the two sides overlap if
	\begin{equation}\label{V0Sin}
	2\mu\ge v\left|\Delta \mathbf{k}_W\right|\;.
	\end{equation}
	An analogous expression holds if $\mu=0$ and $V_0\ne0$. The transmission function is, at all energies, strongly peaked around the Weyl nodes, both in the case of normal and of Klein tunneling, as exemplified for the latter in Fig.~\ref{fig:T1}. In particular, it is nonzero in the area where the projections of the Fermi surfaces from the two sides overlap \cite{Tchoumakov2021}. The transmission probability can reach unit value only for $\theta=0$, and is mildly suppressed for small tilt. 
	\begin{figure}
		\centering
		\includegraphics[width=0.9\columnwidth]{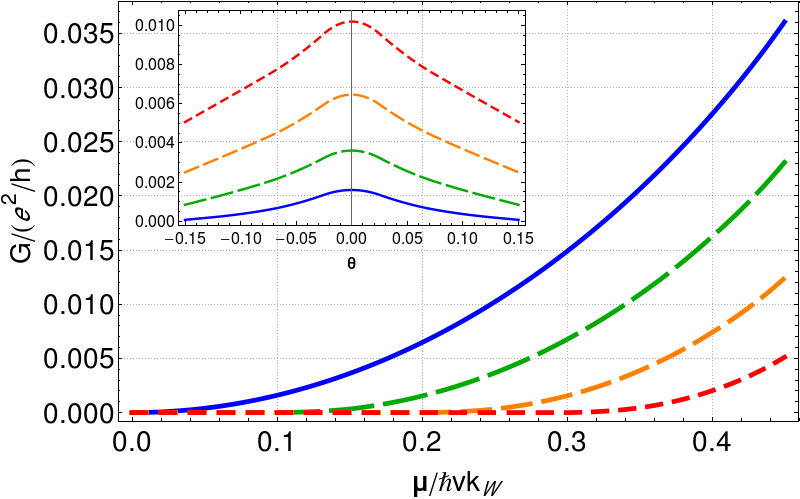}
		\caption{Zero-temperature conductance per unit surface area as a function of chemical potential 
			for $V_0=0$ and (from top to bottom) \mbox{$\theta=0$, $\theta=0.2$, $\theta=0.4$, $\theta=0.6$}. 
			This function is symmetric under $\mu\to-\mu$ in the absence of a potential step. Inset: angular dependence for $\mu=0$ and, from top to bottom, $V_0=0.1vk_W$, $V_0=0.15vk_W$, $V_0=0.2vk_W$, $V_0=0.25vk_W$.}
		\label{fig:conductivity0}
	\end{figure}
	\begin{figure}
		\centering
		\includegraphics[width=0.9\columnwidth]{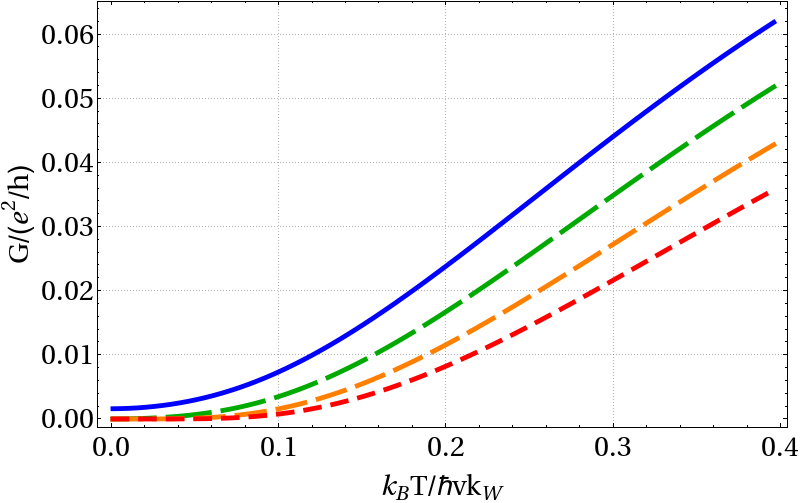}
		\caption{Dimensionless conductance per unit surface as a function of temperature for $V_0=0$, $\mu=0.1vk_W$ and, from top to bottom, \mbox{$\theta=0$}, $\theta=0.2$, $\theta=0.4$, $\theta=0.6$.}
		\label{fig:conductivityT}
	\end{figure}
	
	We now address the consequences of the shape of the transmission factor on transport observables. Throughout our analysis, we assume coherent transport in a clean sample. For $V_0\ne 0$, the charge accumulation in the region around the junction creates a non-uniform electric field, which has been analyzed in \cite{Li2016}. The new features introduced by the tilt are qualitatively similar to that of a normal junction, so we focus on $V_0=0$ in the following.
	Approximating the quasiparticle distributions in the contacted samples with Fermi distributions, we make use of 
	the Landauer-Büttiker formalism to describe quantum transport through the junction. To this end, we define
	\begin{eqnarray}\label{Tthetaint}
	{\cal T}_\theta\left(\varepsilon\right)=\intop \frac{d^2\mathbf{k}}{\left(2\pi \right)^2}\;{\cal T}_\theta\left(\varepsilon,\mathbf{k}\right)\;,
	\end{eqnarray}
	where the integration is over the domain in which there exist incoming states, i.e., in which $k_x$ in \eqref{psiscatt} is real. 
	As exemplified in the inset of Fig. \ref{fig:conductivity0} for various values of the potential step, the total transmission function has a quadratic angular dependence around its maximum at $\theta=0$.
	In terms of the integrals ($k_B=1$)
	\begin{equation}
	I_n=\intop d\varepsilon  \frac{\left(\varepsilon-\mu\right)^n{\cal T}_\theta\left(\varepsilon\right)}{4T\cosh^2\frac{\left(\varepsilon-\mu \right)}{2T}}\;,
	\end{equation}
	we write the charge conductance per unit area as \cite{Li2016}
	\begin{equation}\label{LB}
	G_\theta\left(\mu,T\right)= \frac{e^2}{2\pi} I_0
	\;,
	\end{equation}
	where $e$ is the charge of the electron taken with its sign.
	At low temperatures, this quantity behaves as \mbox{$G_\theta\left(\mu,T\right)\approx G_\theta\left(\mu,0\right)$}, with \mbox{$G_\theta\left(\mu,0\right)=e^2{\cal T}_\theta\left(\mu\right)/{2\pi}$}.
	This function is plotted in Fig.~\ref{fig:conductivity0} for various values of $\theta$. 
	This limit is valid up to quadratic corrections in $T/\mu$: using $\mu=0.1 vk_W$, $v\approx 10^{5}m/s$ and $k_W\approx 9 \times 10^2 \mbox{\AA}^{-1}$ \cite{Lv2015}, one estimates a reference temperature $\mu \approx 70 K$. Clearly, the current is always suppressed around charge neutrality, due to the vanishing density of states.
	Moreover, the momentum-space area where the transmission function is vanishing broadens with increasing tilt angle, 
	due to the mismatch between the scattering states on the respective Fermi surfaces on the two sides \cite{Yang2019}.
	As a consequence, one finds from \eqref{LB} a finite zero-temperature value and a quadratic low-temperature correction proportional to $\frac{e^{2}\pi^2}{48}\partial^2_\mu \mathcal{T}_{\theta}\left(\mu\right)$ only for $\theta=0$, while for small but finite $\theta$ one has activated behavior, with activation gap of the order $\frac{v}{2}\left|\Delta \mathbf{k}_W\right|-\mu$. This is consistent with the temperature dependence of the conductance, shown in Fig.~ \ref{fig:conductivityT} for various tilt angles. For large angles $\theta\approx\pi$, 
	the relevant gap is instead determined by the separation between the nodes with opposite chirality. 
	
\begin{figure}
		\centering
		\includegraphics[width=0.9\columnwidth]{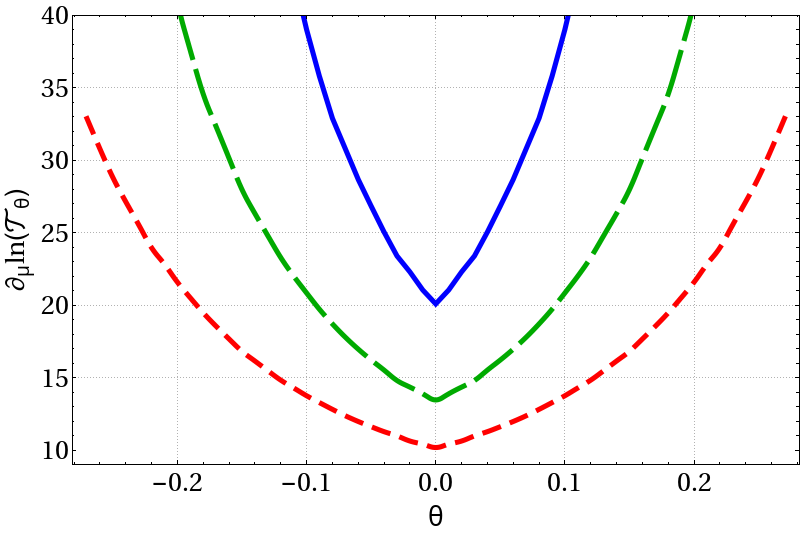}
		\caption{Angle-dependence of the linear contribution \eqref{Slin} to the thermopower for $V_0=0$ and various values of the chemical potential. From top to bottom: $\mu=0.1\,vk_W$, $\mu=0.15\,vk_W$, $\mu=0.2\,vk_W$.}
		\label{fig:dlnT}
	\end{figure}

	A small temperature difference $\Delta T$ between the two contacted samples makes free electrons and holes diffuse across the interface, creating, in the steady state, an electrical voltage gradient $\Delta V_0$. This is known as thermoelectric effect and can be quantified via the thermopower (or Seebeck coefficient) \cite{Benenti2017}
	\begin{equation}\label{S}
	S	=	-\left(\frac{\Delta V_0}{\Delta T}\right)_{j_x=0} = 
	\frac{1}{eT}\frac{I_1}{I_0}\;.
	\end{equation}
	In bulk Dirac and Weyl semimetals, it is known that the numerical value of this quantity is determined by the Berry curvature, the density of carriers and the direction and magnitude of the applied magnetic field, allowing it to achieve very large values \cite{Skinner2018,Das2019}; conversely, in our analysis, the energy-dependence of the  transmission probability across the interface plays a pivotal role.
	It is readily seen that the thermopower is sensitive to the asymmetry of the transmission coefficient: hence, it will be exactly vanishing for $\mu=0$ (it is negative for $\mu>0$ and positive for $\mu<0$). We assume for simplicity $V_0=0$ in the following. At finite chemical potential and low temperature $T\ll\mu$, the first nonzero contribution is linear in temperature and given by the Mott formula
	\begin{equation}\label{Slin}
	S=\frac{\pi^{2}T}{3e}\frac{\partial_\mu{\cal T}_{\theta}\left(\mu\right)}{{\cal T}_{\theta}\left(\mu\right)}\;,
	\end{equation}
	in which the condition \eqref{V0Sin} is assumed for this term to be finite. The coefficient of the linear contribution has a minimum at $\theta=0$ and increases linearly with the tilt angle at small tilts, as shown in Fig.~\ref{fig:dlnT}.
	For large temperatures $\mu\ll k_B T$ the hyperbolic cosine in \eqref{LB} and \eqref{S} flattens and we obtain instead a $1/T$ decay.
	\begin{figure}[h]
		\centering
		\includegraphics[width=0.9\columnwidth]{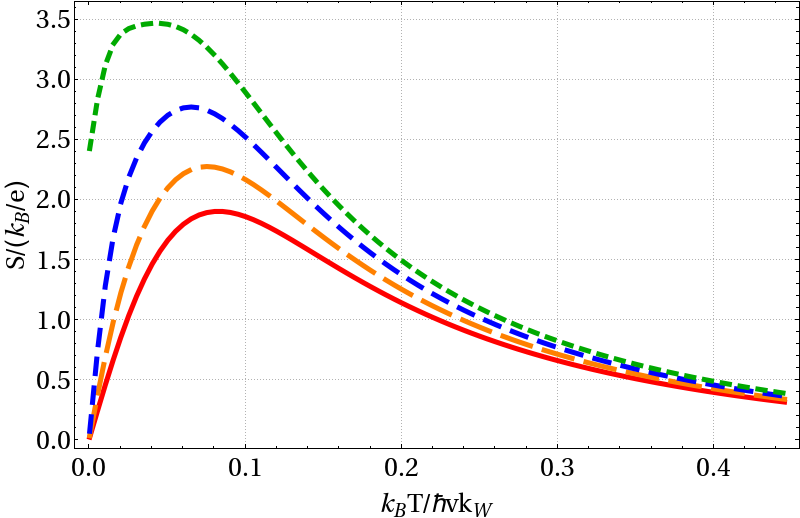}
		\caption{Temperature dependence of the thermopower \eqref{S} at chemical potential \mbox{$\mu=0.15\,vk_W$}, for various tilt angles. From bottom to top: $\theta=0$,  $\theta=0.1$, $\theta=0.2$ and $\theta=0.3$.}
		\label{fig:thermoT}
	\end{figure}
	As seen above, the tilt angle strongly suppresses the charge conductance, which has the consequence that \eqref{S} must be larger for larger tilt angles, as shown in Fig.~\ref{fig:thermoT}, and for lower values of the chemical potential. We observe that the position of the maximum of the thermopower is shifted toward lower and lower temperatures as $\mu$ is decreased. In fact, for finite chemical potential, the particle branch has a larger density of states than the hole branch and the numerator of \eqref{S} initially increases. However, at larger energies, more states in the hole branch are accessible, contributing with the opposite sign to the thermoelectric current and the thermopower starts decreasing. 
	
	Summing up, we have shown that the electric conduction and the thermoelectric properties of the junction can be tuned by the tilt angle $\theta$. The considerations above only arise from the probability of the electron being transmitted at the interface, independently from its direction. However, 
	we note that the momentum in the $x$ direction is generally discontinuous across the interface: this leads to electron refraction at the interface, which we study in detail in the next section.

	\section{Refraction at the interface}\label{sec:Refraction}
	
	Given that the dispersion relation is anisotropic, an electron 
	with the same momentum has a different energy on the two sides of the interface. 
	In particular, as translation invariance is broken only in the $x$ direction, 
	the component of the momentum parallel to the interface is continuous across it, 
	while $k_x$ is discontinuous and determined from \eqref{kx} and \eqref{kthetax}. 
	Considering a monochromatic electron beam incident from a given direction, we determine how the direction of the transmitted excitation depends on $\theta$ and $V_0$, both in the cases of normal and Klein tunneling. Complementing earlier observations \cite{Yang2019}, we show that the refraction angle is not uniquely determined by the incidence angle, but one has to specify the chirality of the incident electron as well. We explore the effect of the tilt on the polar angle and determine the splitting of the electron beam due to the chirality of the nodes.
	
\begin{figure}
		\centering
		\includegraphics[width=0.9\columnwidth]{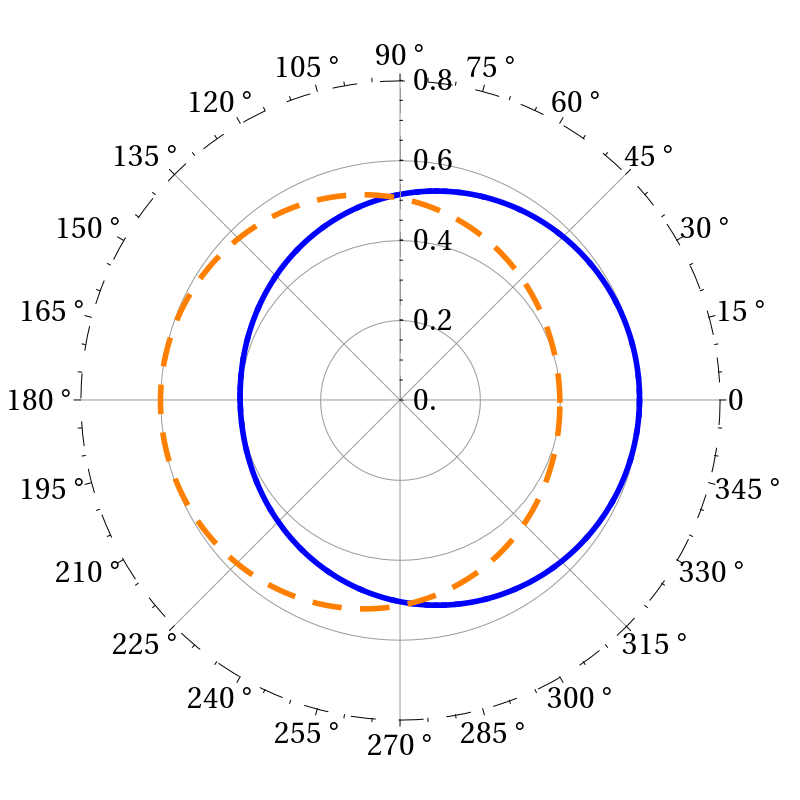}
		\caption{Polar plot of the refraction polar angle for Klein transmission, see Eq.~\eqref{coschio}. 
			We represent above $\sin\chi_o$ as a function of $\xi_i$ at fixed incidence angle $\chi_i=0.1$ for $V_0=0.2\,vk_W$, $E=0$, and $\theta=0.1$.
			The angle $\chi_o$ depends on the chirality of the incident electron:
			the continuous blue line is for an incident electron with positive chirality $\eta=1$, 
			the dashed orange line for negative chirality $\eta=-1$. 
			The two curves are exchanged if the tilt angle is reversed ($\theta\to-\theta$)}.
		\label{fig:circleshole}
	\end{figure}

	As seen in Sec.~\ref{sec:Transmission}, the transmission of an electron takes place in the vicinity of a Weyl node. As a consequence, its dispersion relation is approximately linear in the deviation of the momentum from the position of the Weyl node and the group velocity of the incoming electron is
	\begin{equation}\label{vi}
	\left(v_{i,x},\mathbf{v}_i \right)
	=\frac{v^2}{\varepsilon_+}\left(k_x, \mathbf{k} - \mathbf{k}^{(\eta)}_W\right)\,,
	\end{equation}
	where $\eta=\pm$ and $\varepsilon_\pm=E\pm V_0$. The momentum $\mathbf{k}$ parallel to the interface is unchanged through the interface, but the Weyl nodes are in position $\mathbf{k}_{W,\theta}=\bar{\eta}k_{W}\left(-\sin\theta,\cos\theta\right)$, where $\bar{\eta}$ is now the chirality of the Weyl node in which the electron is transmitted, i.e., $\bar{\eta}=\eta$ for $\theta\sim0$ and $\bar{\eta}=-\eta$ for $\theta\sim\pi$. It follows that the velocity of the outgoing mode is
	\begin{equation}\label{vout}
	\left(v_{o,x},\mathbf{v}_{o}\right)
	=\frac{v^2}{\varepsilon_-}\left(\bar k_{x},\mathbf{k}-{\bf k}_{W,\theta}^{(\bar\eta)} \right)\,,
	\end{equation}
	with ${\bf k}_{W,\theta}^{(\bar\eta)}$ defined in \eqref{rotkW}.
	Importantly, the sign of $\varepsilon_-$ can be negative, which signals Klein tunneling, in that the velocity and momentum of a hole are opposite in direction.
	We denote the angle of the incident/outgoing particle velocity with respect to the normal to the interface as 
	$\chi_{i,o}$ and the azimuthal angle in the $yz$ plane as $\xi_{i,o}$, so that
	\begin{equation}\label{velparamet}
	(v_{i/o,x},\mathbf{v}_{i/o})=v(\cos\chi_{i/o},\sin\chi_{i/o}\cos\xi_{i/o},\sin\chi_{i/o}\sin\xi_{i/o}),
	\end{equation}
	with $0\le \chi_{i/o} < \pi/2$ and $0\le \xi_{i/o} < 2\pi$.
	The momentum component perpendicular to the interface of a transmitted excitation is a function of the energy and the momentum parallel to the interface, which are conserved in the transmission process. 
	With the parameterization \eqref{velparamet}, we find
	\begin{eqnarray}
	v^2 \bar{k}_{x}^2 &=&\varepsilon_{-}^{2}-\varepsilon_{+}^{2}\sin^{2}\chi_{i}-
	2v^{2}k_{W}^{2}\left(1-\eta\bar{\eta}\cos\theta\right)\\
	&&-2\eta vk_{W}\varepsilon_{+}\sin\chi_{i}\left[\sin\xi_{i}-\eta \bar{\eta}\sin\left(\xi_{i}-\theta\right)\right]\nonumber \;.
	\end{eqnarray}
	From Eqs.~\eqref{vout} and \eqref{velparamet}, the polar angle is then given by
	\begin{equation}
	\cos\chi_{o}=\frac{v\bar{k}_{x}}{\left|\varepsilon_-\right|}. \label{coschio}
	\end{equation}
	Moreover, setting $\xi_o=\xi_i+\phi$ for an outgoing electron (or $\xi_o=\xi_i+\phi+\pi$ for an outgoing hole), we find 
	\begin{equation}
	\tan \phi = \frac{-2\eta vk_W\sin \frac{\theta}{2} \sin (\xi_i- \frac{\theta}{2}) }{|\varepsilon_+|\sin\chi_i+2\eta vk_W \sin\frac{\theta}{2}\cos(\xi_i-\frac{\theta}{2})}
	\label{phi}
	\end{equation}
	if the transmission takes place close to a node of the same chirality, i.e., if $\eta\bar \eta=1$. If the chirality is changed across the interface, for $\eta\bar \eta=-1$, one should replace $\theta$ with $\theta+\pi$ in Eq.~\eqref{phi}. 
	The equations \eqref{coschio} and \eqref{phi} fix the direction of the transmitted electron or hole 
	and can be seen as a generalization of Snell's law \cite{Hills2017}. As an example, Fig.~\ref{fig:circleshole} illustrates the dependence of the polar angle of a transmitted hole on the azimuthal angle of the incident electron.
	
\begin{figure}
		\centering
		\includegraphics[width=0.9\columnwidth]{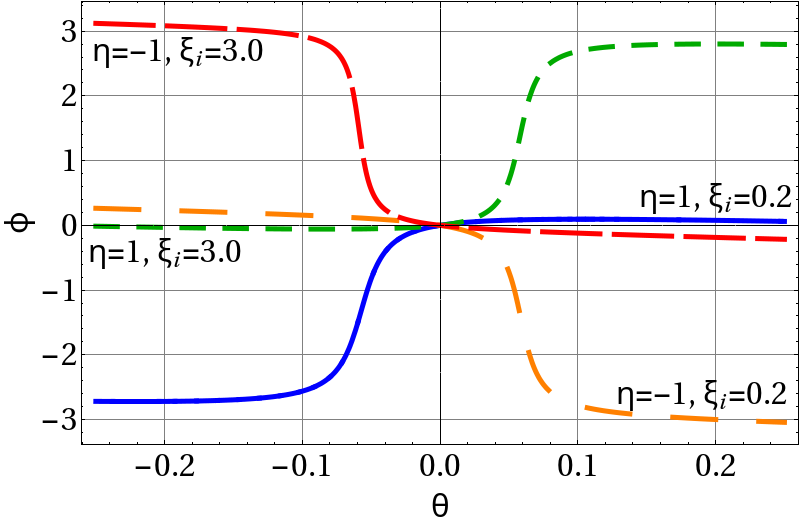}
		\caption{Azimuthal angle shift for a transmitted electron with $E=0.2\,vk_W$ and $V_0=0$. We represent above $\phi$ as function of $\theta$ for two fixed incidence angles $\chi_i=0.3$, $\xi_i=0.2$ and  $\chi_i=0.3$, $\xi_i=3.0$, and the two chiralities.}
		\label{fig:xi}
	\end{figure}

	The presence of a tilt already implies that the angle of refraction is different from the angle of incidence, and the anisotropy of the material makes the refraction coefficient dependent on the azimuthal angle $\xi_i$. 
	Noticeably, normal incidence ($\chi_i=0$) does not imply normal transmission, but instead transmission at the angle
	\begin{equation}\label{sinchio}
	\sin\chi_o=
	\frac{2vk_W}{\left|\varepsilon_-\right|}\sin\frac{\theta}{2}
	\end{equation}
	with respect to the normal.
	In order to underline the effect of the tilt on the azimuthal angle, one can consider a situation in which the component of the momentum of the transmitted electron parallel to the separation between the Weyl nodes on the two sides of the interface lies between the projection of the two Weyl nodes. Then, the projection of the velocity in this direction is opposite on the two sides of the junction, which results in a large shift of the angle $\phi$. This mechanism is exemplified in Fig.~ \ref{fig:xi}, in which the momentum of the electron is held fixed, i.e., it is a point in the Brillouin zone: by increasing $\theta$, the Weyl node passes from one side to the other of this point, hence, the azimuthal angle shift $\phi$ quickly passes from $\approx0$ to $\approx\pi$ when this happens.
	
	There is an explicit dependence of the refraction angles in \eqref{coschio} and \eqref{phi} on the chirality $\eta$ of the incident particle: the fact that the same group velocity is attained by electrons near both Weyl nodes implies that birefraction takes place and the interface acts like a beam splitter for the electrons. As shown in Fig.~\ref{fig:circleshole}, the refraction angle depends on the azimuthal angle and a \emph{finite} tilt angle displaces the electrons with the same incidence direction, but opposite chiralities, in opposite directions. 
	The splitting angle $\chi_s$ between the refracted beams can be directly computed 
	using the angle parametrization \eqref{vout}. For $|\theta|<\pi/2$, using Eqs.~\eqref{coschio} and \eqref{phi}, we arrive at
	\begin{widetext}
		\begin{equation}
		\cos\chi_s 
		=
		\frac{v^2k_W^2}{\varepsilon_-^2}\left\lbrace    \frac{\varepsilon_+^2\sin^2\chi_i}{v^2k_W^2}-4\sin^2\frac{\theta}{2}+\sqrt{    \left[\frac{\varepsilon_-^2}{v^2k_W^2}-4\sin^2\frac{\theta}{2}-\frac{\varepsilon_+^2\sin^2\chi_i}{v^2k_W^2}\right]^2-16\frac{\varepsilon_+^2}{v^2k_W^2} \sin^2\chi_i \sin^2\frac{\theta}{2}\cos^2\left(\xi_i-\frac{\theta}{2}\right)    }    \right\rbrace
		.
		\end{equation}
	\end{widetext}
	
\begin{figure}[h]
		\centering
		\includegraphics[width=0.8\columnwidth]{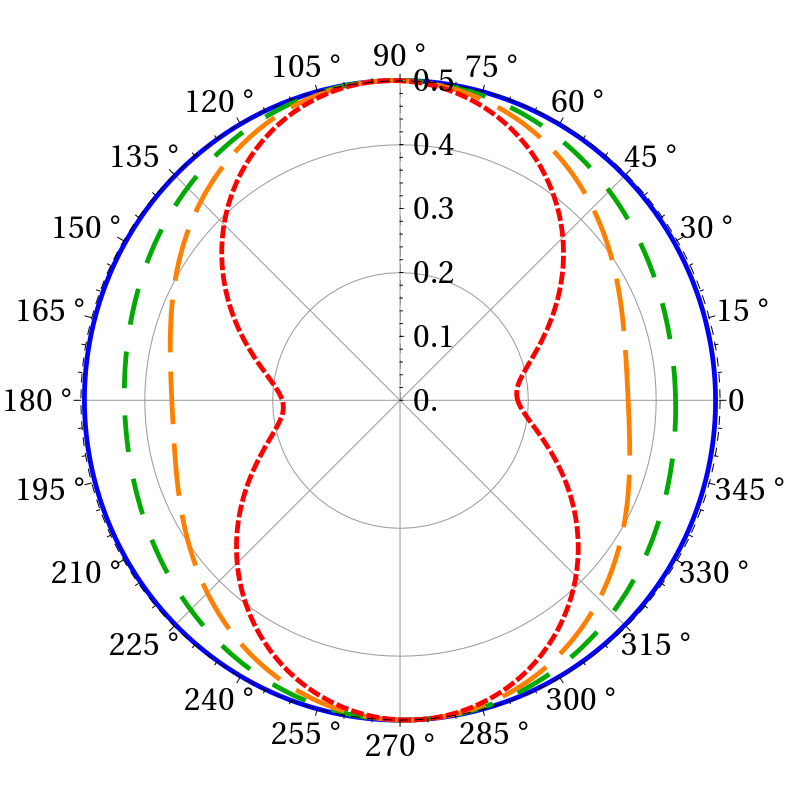}
		\caption{Polar plot of the splitting angle $\cos\chi_s$ as a function of the azimuthal angle $\xi_i$ for $E=0$, $\theta=0.1$ and $V_0=0.2vk_W$, for various incidence angles: starting from the outer curve and proceeding inwards $\chi_i=0.1$, $\chi_i=0.3$, $\chi_i=0.4$ and $\chi_i=0.5$.}
		\label{fig:split}
	\end{figure}

	As illustrated in Fig.~\ref{fig:split}, the splitting angle is minimal for \mbox{$\xi_i=\left(\theta\pm\pi\right)/2$}, while it reaches its maximum value in the orthogonal direction \mbox{$\xi_i=\theta/2$}, \mbox{$\xi_i=\pi +\theta/2$}. In the case of normal incidence $\chi_i=0$, the \mbox{splitting angle} is independent of $\xi_i$.
	For Klein tunneling in a Weyl $np$-junction, as well as for normal tunneling at specific angles, the transmitted particle is refracted with opposite component of the velocity in the plane parallel to the interface. Therefore, the possibility of focusing electron beams, or Veselago lensing \cite{Cheianov2007,Chen2021,Hills2017}, appears, but it is hindered by the intrinsic anisotropy of the materials if the Weyl nodes on the two sides of the interface are misaligned. 
	On the other hand, the interface acts as a polarizing filter, in which the "polarization" is the chirality index, due to the fact that a monochromatic electron beam incoming from a given direction is split at the junction. For materials with more Weyl nodes, our analysis above suggests a different outgoing angle for each of the nodes.

	\section{Interface states}\label{sec:Interfacestates}
	
	We now study states which are exponentially localized at the interface. 
	These states are expected, for instance, in related systems locally described 
	in the bulk by a Dirac equation, and at the interfaces between topological 
	insulators \cite{Shen2011,Sen2012}. 
	Defining the inverse decay lengths
	\begin{eqnarray}
	\kappa & =\frac{1}{v}\sqrt{v^2k_y^2+m^2(k_z) - \varepsilon_+^2} \;,\qquad &x<0\,, \\
	\bar\kappa & =\frac{1}{v}\sqrt{v^2k_{\theta,y}^2+m^2(k_{\theta,z}) - \varepsilon_-^2} \;,\qquad &x>0\,,
	\end{eqnarray}
	a localized eigenstate of the Hamiltonian \eqref{Hint} can be written in the form
	\begin{equation}\label{interfacestates}
	\psi_{E,\mathbf{k}}(x)	= \begin{cases}
	C_1 u_{0;-i\kappa,\mathbf{k}} e^{\kappa x} & x<0\\
	C_2 u_{\theta;i\bar \kappa,\mathbf{k}} e^{-\bar \kappa x}
	& x>0
	\end{cases}\;,
	\end{equation}
	where we use the spinors defined in \eqref{bulkeigenvec}. $C_1$ and 
	$C_2$ are arbitrary coefficients, determined by continuity and normalization of the wave function.
	Imposing the continuity at the interface, we arrive at the condition
	\begin{equation}\label{interfacecontinuity}
	\Phi\left(E,\mathbf{k}\right)=1\;,
	\end{equation}
	where
	\begin{equation}\label{Phifunction}
	\Phi\left(E,\mathbf{k}\right) = \frac{\left(E+V_{0}-m\left(k_{z}\right)\right)\left(-v\bar \kappa +b_{\theta,y}\left(\mathbf{k}\right) \right)}{\left( E-V_{0}-b_{\theta,z}\left(\mathbf{k}\right)\right)
		\left( v\kappa+vk_{y}\right)}.
	\end{equation}
	This equation implicitly defines the dispersion relation $E=E(\mathbf{k})$ of the interface states.
	For fixed energy, its solutions define a one-dimensional curve in the Brillouin zone, 
	an {\em interface Fermi arc}, analogous to a surface Fermi arc. 
	A necessary condition for such interface states to exist is that $\kappa$ and $\bar \kappa$ must be real. The endpoints of the arc are determined by the conditions $\kappa=0$ or $\bar \kappa=0$: in these points, the interface states merge with the bulk states on one or the other side. Interface arcs exist for \mbox{$|E|>V_0$} for every $\theta>0$, while they exist for \mbox{$|E|<V_0$} if the projected Fermi surfaces do not overlap.
	
	\begin{widetext}
		
		\begin{figure}
			\centering
			\includegraphics[height=0.185\textheight]{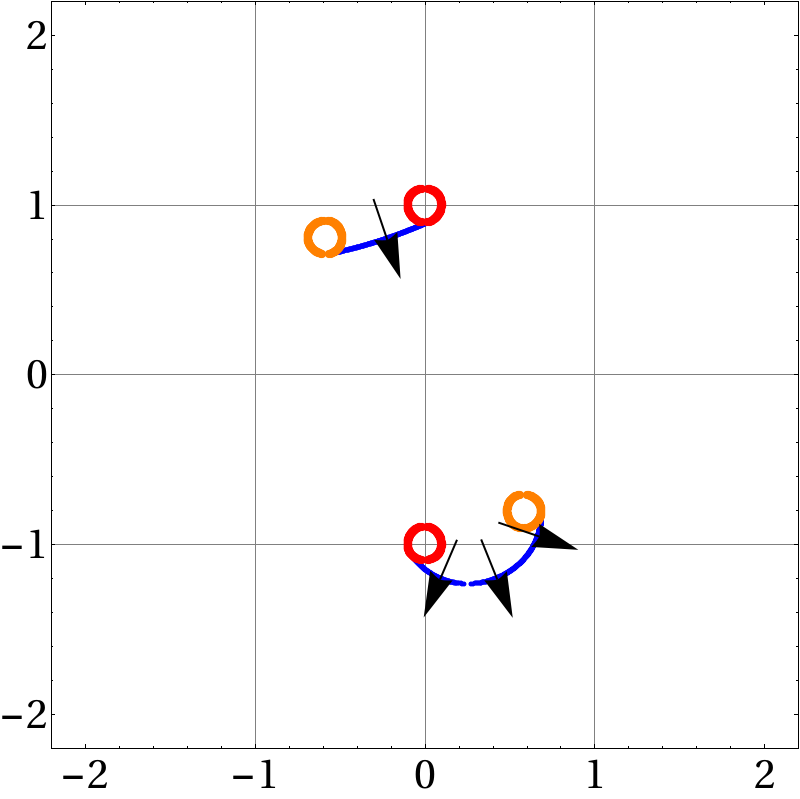}
			\includegraphics[height=0.185\textheight]{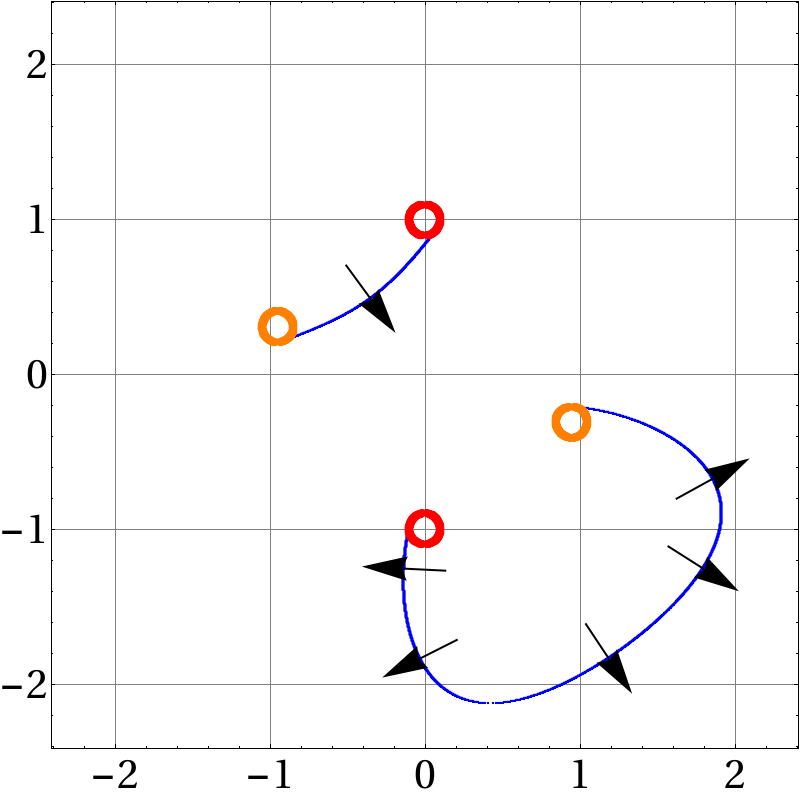}
			\includegraphics[height=0.185\textheight]{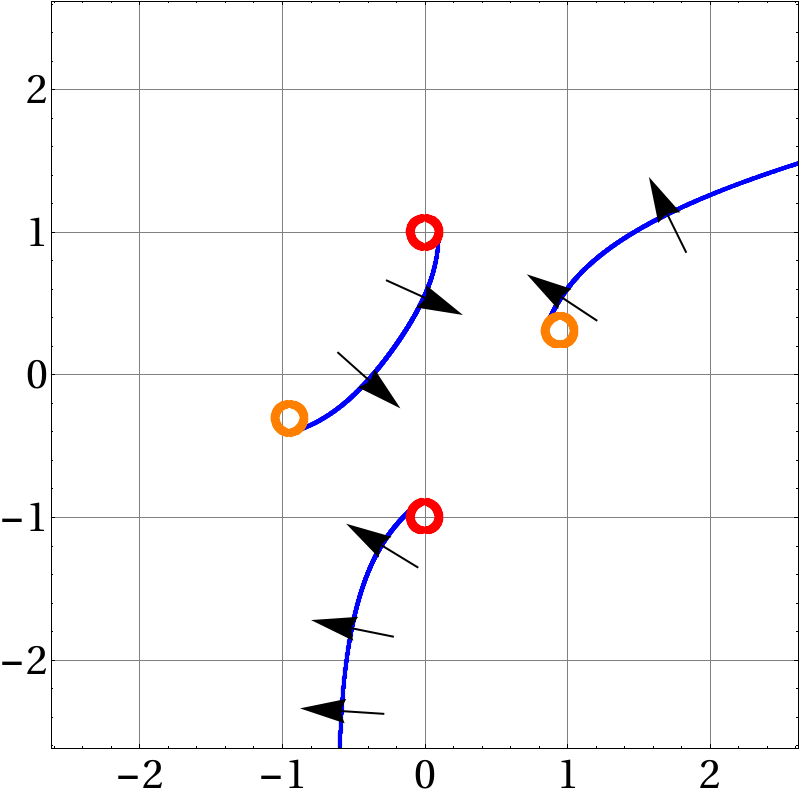}
			\includegraphics[height=0.185\textheight]{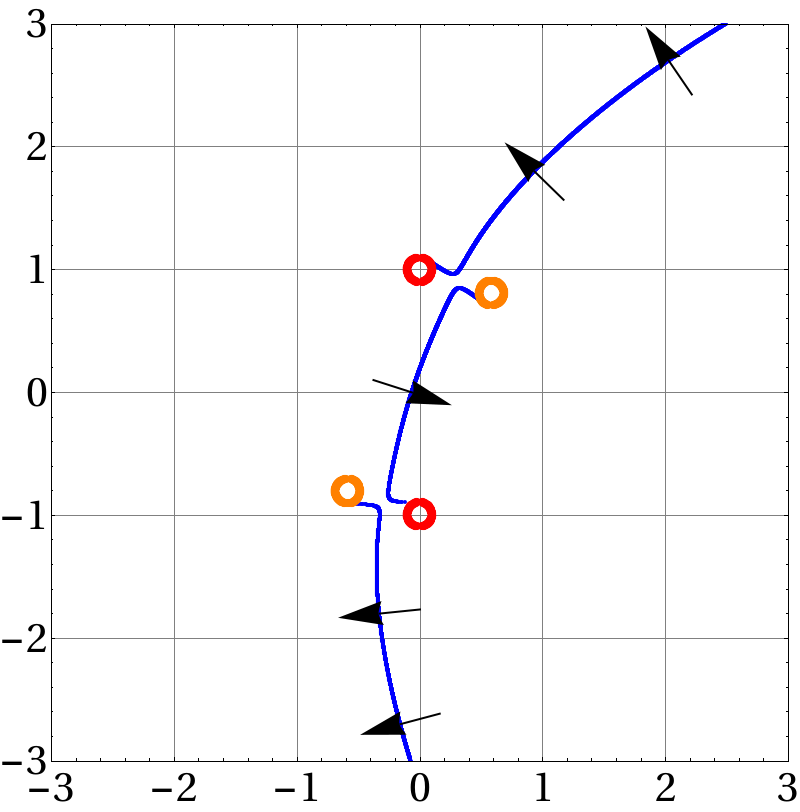}
			\caption{Interface states in the plane $k_y-k_z$ (in units of $k_W$) for \mbox{$V_0=0$}, energy $E=0.1\,vk_W$, 
				and several values of the tilt angle $\theta$: from left to right, $\theta=\pi/5$, $2\pi/5$, $3\pi/5$, $4\pi/5$.
				The circles correspond to the interface projections of the bulk Fermi surfaces, 
				the red being those of the left subsystem, with Weyl nodes at $(0,\pm k_W)$,
				the orange the ones of the right, rotated subsystem, with Weyl nodes 
				in the positions \eqref{rotkW}. The black arrows represent the direction 
				of the group velocity, which is normal to the arc. 
				In the last two panels, the incomplete arc portions join outside the shown region of the
				$k_y-k_z$ plane.}
			\label{fig:shape0}
		\end{figure}
		
		\begin{figure}
			\centering
			\includegraphics[height=0.185\textheight]{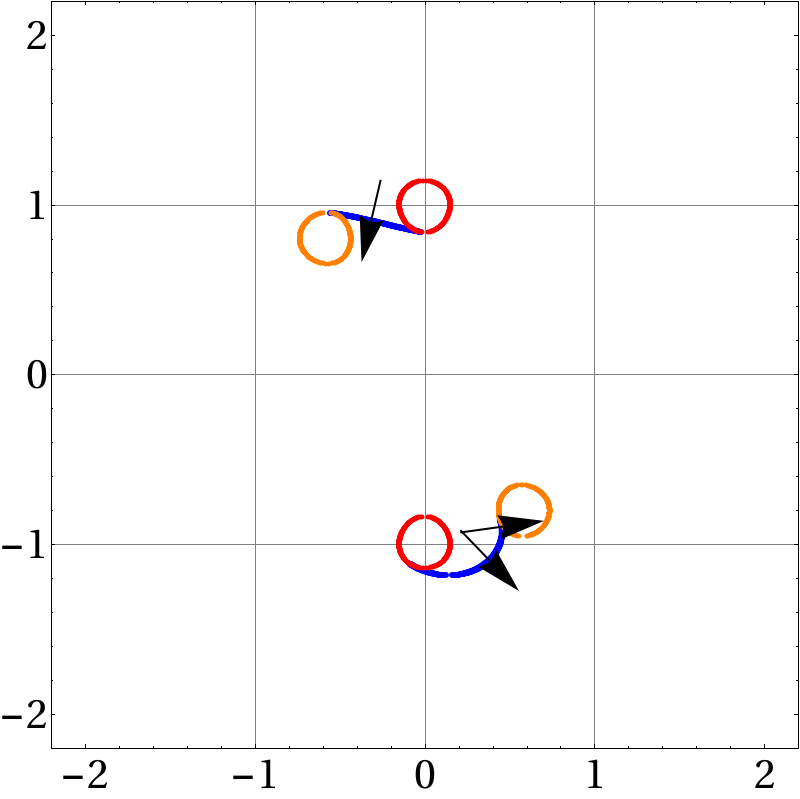}
			\includegraphics[height=0.185\textheight]{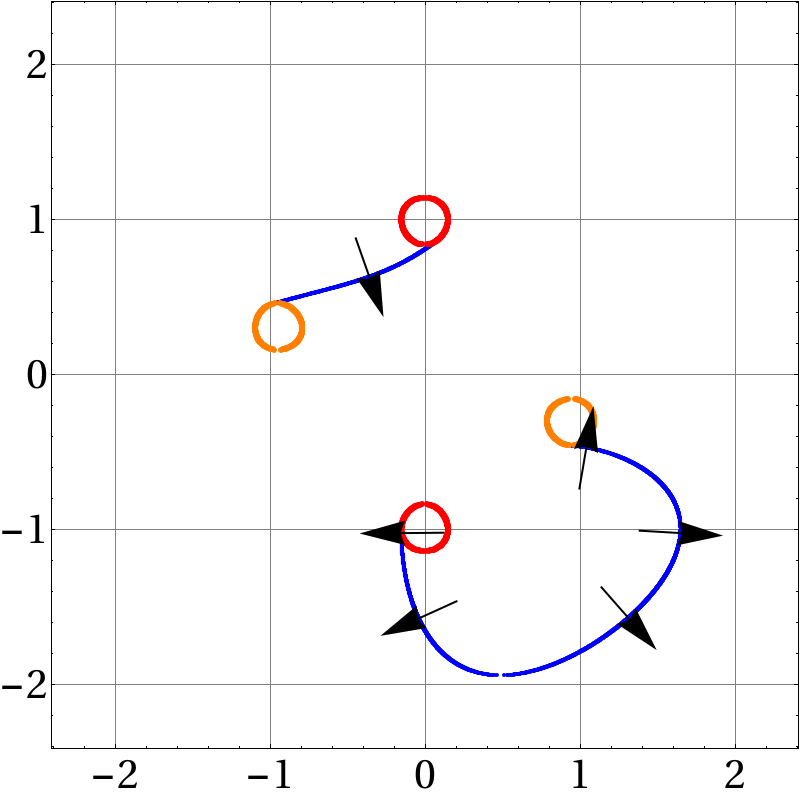}
			\includegraphics[height=0.185\textheight]{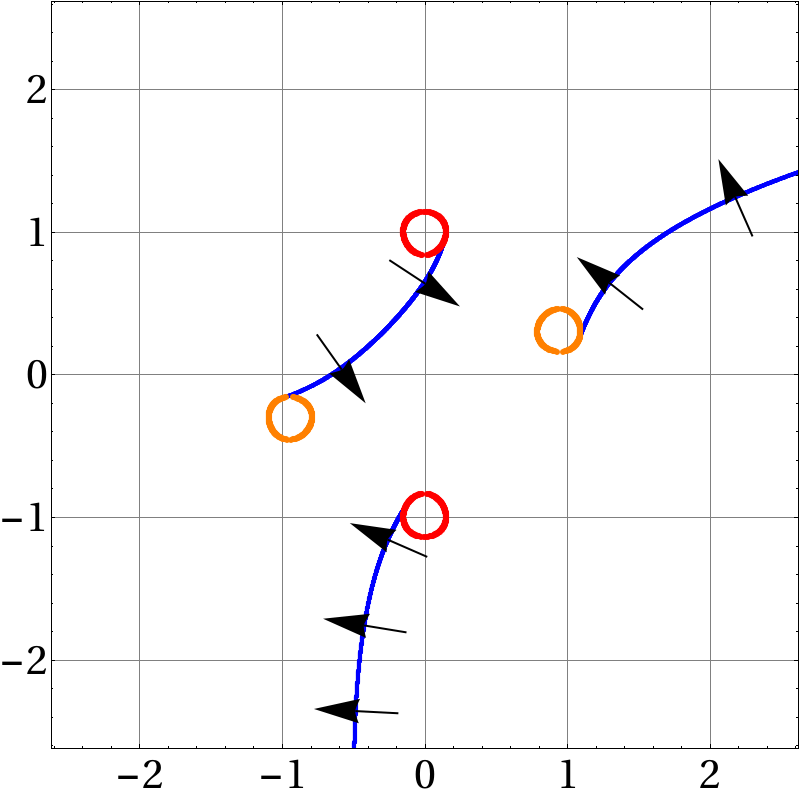}
			\includegraphics[height=0.185\textheight]{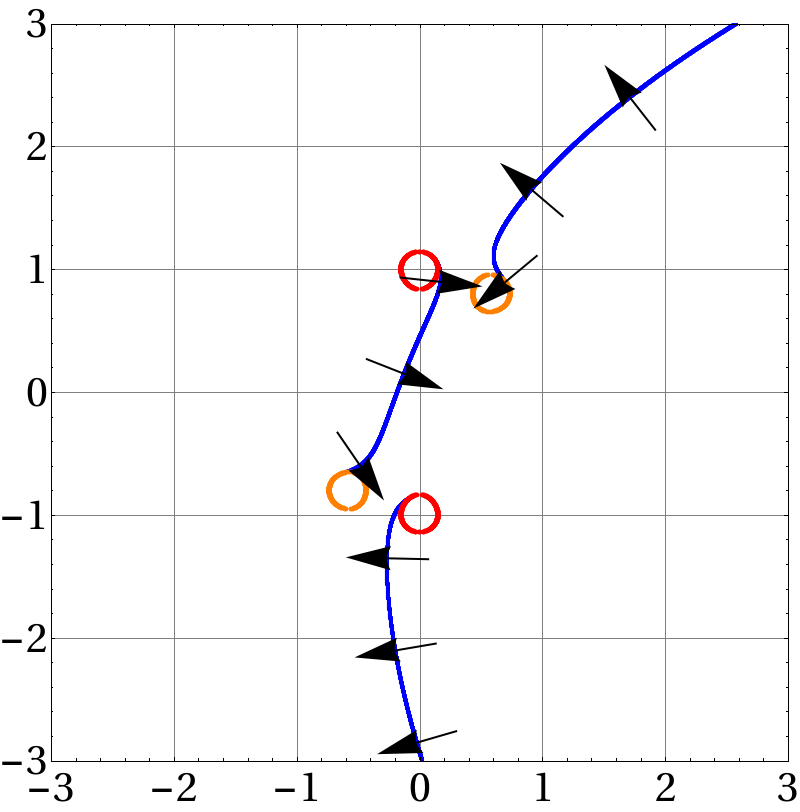}
			\caption{Zero-energy interface states in the plane $k_y-k_z$ (in units of $k_W$) 
				for a potential step with \mbox{$V_0=0.15 vk_W$} and the same values of $\theta$ 
				as in Fig.~\ref{fig:shape0}. The conventions are the same as in Fig.~\ref{fig:shape0}.
				In this case no interface state is present as long as the bulk Fermi surface projections overlap. As soon as they are disconnected, 
				an interface state connects the projections with the same chirality. 
			}
			\label{fig:shapeNP}
		\end{figure}
		
	\end{widetext}
	Using the implicit function defined by \eqref{interfacecontinuity}, 
	one can also compute the group velocity of an electron wave packet on the interface as
	\begin{equation}
	v_a=-\frac{\partial_{k_a}\Phi }{\partial_E\Phi},\qquad\quad a=y,z \:.
	\end{equation}
	The shape of the interface arcs can be determined by solving \eqref{interfacecontinuity} 
	numerically, and is illustrated in Figs.~\ref{fig:shape0} and \ref{fig:shapeNP} for 
	several values of the tilt angle in two different situations. 
	In Fig.~\ref{fig:shape0} we show the interface arcs for the case $V_0=0$, while 
	Fig.~\ref{fig:shapeNP} shows the zero-energy arcs for the case of finite $V_0$ ($np$ junction setup).
	The first interesting feature we see is that, 
	in contrast to the usual surface Fermi arcs that connect (the projections of) 
	Weyl nodes of opposite chirality, the interface arcs connect the projections of the bulk Fermi surfaces of opposite
	sides of the interface with the {\em same chirality}. 
	This occurs for any value of $\theta$ for which interface arcs
	exist. It is interesting to note that in the case $V_0=0$, 
	below a critical angle $\theta_c$, the shorter arc connects the
	nodes with chirality $\eta=+1$, the longer arc those with chirality $\eta=-1$. 
	
	At $\theta=\theta_c$ the two arcs intersect and then exchange their role, 
	as seen in the last two panels of Fig.~\ref{fig:shape0}.
	For example, in our model, at $E=0.1vk_W$, we find $\theta_c\approx 2.49$, close to $4\pi/5$, as can be seen in the last panel of Fig.~\ref{fig:shape0}. 
	The angle $\theta_c$ tends to $\pi$ as $E\to0$, but otherwise its value is model-dependent.
	If the Weyl node separation is not the same 
	on the two sides, there appears a range of angles 
	in which the connectivity changes, namely, the arcs connect the Weyl nodes 
	of opposite chirality on the same side. 
	When the separations become the same, this 
	interval shrinks to zero, and the change of connectivity occurs only at $\theta=\pi$.
	This is in accordance with the results of \cite{Ishida2018}. In Fig.~\ref{fig:shape0} we also note that for $V_0=0$ the arcs are symmetric under reflection in the line
	going through the midpoints between the nodes with the same chirality. This is a consequence of the 
	invariance of Eq.~\eqref{interfacecontinuity} under a reflection symmetry 
	as discussed in App.~\ref{sec:interapp}, and implies that both arcs carry a net current in the direction $(\sin\frac{\theta}{2},-\cos\frac{\theta}{2})$, perpendicular to the displacement vector between the Weyl nodes on opposite sides of the interface $\Delta \mathbf{k}_W $ in \eqref{DeltakW}.
	
	The second salient feature in Figs.~\ref{fig:shape0} and \ref{fig:shapeNP}
	is that the way the arcs attach to the bulk projections depends
	on whether the projections consist of particle states or of hole states. 
	This difference can be rationalized along the lines of \cite{Haldane2014,Armitage2018}. 
	The Fermi contours depicted in the figures are constant-energy curves, so the group velocity, 
	which is the energy gradient in the $k_y-k_z$~plane, is always normal to the curves. 
	At the junction between the arc and the bulk part of the 
	Fermi surface, the velocities of the interface and of the bulk states must eventually align. 
	Looking at Fig.~\ref{fig:shape0}, one sees that this is indeed the case, 
	as the velocity of bulk states is oriented perpendicularly to the circles, 
	pointing outwards. Conversely, the velocity of holes states has the opposite sign: 
	in the presence of a potential step, the arc must therefore  
	attach to the circle on the opposite side, which is what we observe when comparing 
	Fig.~\ref{fig:shapeNP} with Fig.~\ref{fig:shape0}.

	The detailed shape of the interface arcs depends on the specific form of the model Hamiltonian. However, we can consider the net chirality of the interface states, defined as the difference $\mathcal{N}=n_+-n_-$ between the numbers of right ($n_+$) and left ($n_-$) movers in the $y$ direction at given $k_z$.
	Its change when crossing the projection of the Weyl nodes is independent of the microscopic details and fixed by the relative position of the projections of the Weyl nodes in the bulk subsystems. 
	We can understand the qualitative aspects of the arc shapes in terms of this difference.
	In order to see this, we follow the arguments of \cite{Burkov2011,Ishida2018,Dwivedi2018} and
	divide the Brillouin zone of the system into slices 
	with fixed $k_z$: away from the Weyl nodes, each slice can be seen as the 
	Brillouin zone of the junction between two-dimensional topological insulators.
	As such, the value of $\mathcal{N}$ is fixed by the difference between the bulk Chern numbers \cite{Thouless1982,Qi2006,Hasan2010,Takahashi2011,Mong2011}.
	The continuum Hamiltonian \eqref{Hint} does not have a Brillouin zone, yet, it can be seen as the small-momentum expansion of a lattice model and the role of the difference between the Chern numbers is taken on by the difference of the signs of the mass functions at given $k_z$, i.e. 
	$\mathcal{N}=\mbox{sgn}\left(m\left(k_{\theta,z}\right)\right)-\mbox{sgn}\left(m\left(k_{z}\right)\right)$.
	As we scan in $k_z$, when the mass changes sign across a Weyl node, the number of interface modes changes. In our system, because of the tilt, 
	the sign change takes place at different values of $k_z$ in the right and left subsystems.
	Let us illustrate this argument with the help of, e.g., the second panel of Fig.~\ref{fig:shape0}.
	When $k_z$ crosses the Weyl node of the left subsystem (at $k_z=k_W$), 
	the interface is between a topological insulator and a trivial insulator, 
	so a Fermi arc should appear: 
	indeed, starting from the region $k_z>k_W$, where ${\cal N}=0$, 
	we observe that as soon as $k_z=k_W$, a right mover appears and ${\cal N}$ jumps to $1$.
	The situation is mirrored for negative $k_z$: for \mbox{$k_z<-k_W$}, there are one left and one right interface 
	modes and $\mathcal{N}=0$, while we observe $\mathcal{N}=1$ for $-k_W<k_z<-k_W\cos\theta$.
	As soon as $k_z$ crosses the Weyl node of the right subsystem, 
	we have a junction of two topological insulators with the same value of the Chern numbers, and indeed we observe $\mathcal{N}=0$ in the region \mbox{$|k_z|<k_W\cos\theta$}. 
	If $\theta>\pi/2$, as in the third panel of Fig.~\ref{fig:shape0}, we observe a similar situation, 
	with $\mathcal{N}=0$ if $|k_z|>k_W$, and $\mathcal{N}=1$ if \mbox{$k_W>|k_z|>k_W |\cos\theta|$}. 
	Contrarily to the previous case, in the region \mbox{$|k_z|<k_W |\cos\theta|$} the sign the mass function jumps from $-1$ to $+1$ across the interface and we observe indeed $\mathcal{N}=2$.
	
	
	
	To close this section, we remark that we obtained the arcs for a \emph{transparent} interface, 
	whereas the conventional Fermi arc surface states are found imposing a vanishing condition on the current across the interface.
	Our approach can be seen as the limit in which Fermi arcs from two disconnected samples are fully hybridized by a very strong tunneling between the samples \cite{Dwivedi2018,Murthy2020,Abdulla2021}. From this perspective, the region in which no interface states are present results from the gapping out of counter-propagating Fermi arcs, while this does not happen if the two Fermi arcs are co-propagating, in the region with $\mathcal{N}=2$.
	
	It is worth emphasizing that interactions may alter the transport properties of the interface in the presence of electrons localized at the interface. While this would change the boundary conditions for weakly tunnel-coupled surfaces, it can be neglected in first approximation in our strong tunneling limit.
	An intriguing consequence of the existence of interface arcs can instead
	be observed in the electric transport in the $y$ direction, 
	which we study in a slab geometry in the next section.

	\section{Scattering in the presence of surface states\label{sec:Slabs}}
	
	We now consider the junction of two slabs, with transverse size $L$ in the $y$-direction, 
	but otherwise infinitely extended and joined at $x=0$ via a transparent interface. 
	In this situation, two Fermi arcs are present on the lateral surfaces at $y=0$ and $y=L$ 
	and transport chiral current in the $x$ direction. These states are responsible for the large surface 
	contributions to electric transport in different geometries and setups
	\cite{Baireuther2016,Breitkreiz2019,Pal2021,DeMartino2021}. The transmission of 
	the surface current carried by these states at the interface depends on the twist angle $\theta$. 
	Considering for definiteness a tilt $|\theta|<\pi/2$, the projections of the Weyl nodes on the surface Brillouin zone at $y=0$ for the left subsystem are at distance $2k_W$, while at distance $2k_W\cos\theta$ for the right subsystem. It follows that for $k_W\cos\theta<|k_z|<k_W$, a Fermi arc is present for $x<0$, but not for $x>0$. In this region, as seen in the previous subsection, an interface arc appears, which can indeed be thought of as the continuation of a portion of the Fermi arc in the bulk of this system. The same argument applies for $|\theta|>\pi/2$, only in this case the Fermi arcs at $y=0$ have opposite velocities in the $x$ direction, hence, we have $\mathcal{N}=2$ in the region $|k_z|<-k_W\cos\theta$.
	In order to illustrate the physical consequences,
	we consider here the two extreme cases, $\theta=0$ and $\theta=\pi$, 
	in which the Weyl nodes are at $\mathbf{k}=(0,\pm k_W)$ on both sides of the interface. 
	For the sake of simplicity, we assume straight Fermi arcs on the surfaces of the slab.

	\subsection{Bike lanes}
	For $\theta=0$, we impose the boundary conditions \cite{Witten2015}
	\begin{eqnarray}\label{bc+-}
	\sigma^{x}\Psi\left(y=0\right)&=&\Psi\left(y=0\right) \nonumber \\
	\sigma^{x}\Psi\left(y=L\right)&=&-\Psi\left(y=L\right)
	\end{eqnarray}
	on the wave function $\Psi$, for every value of $x$. These boundary conditions do not break inversion symmetry
	\cite{Buccheri2021} and imply that the charge current vanishes across the surfaces at $y=0$ and $y=L$. 
	The wave function reduces then to the eigenvectors $\xi_\pm$ of the Pauli matrix $\sigma^x$. 
	
	For simplicity, we consider the large-$L$ limit. 
	The spectrum is then composed of bulk states in the form  \eqref{bulkstates0}
	(plane waves in the $x$- and $y$-direction). 
	The transverse momentum is quantized according to Eq.~\eqref{bulkquanteq}. 
	In the large transverse size limit \mbox{$|m|L\gg 1$}, the quantized $k_{y,n}$ 
	approach the values
	\begin{equation}\label{kyn}
	k_{y,n}=\frac{\pi n}{L}, \quad n=1,2,\dots.
	\end{equation}
	The corresponding transverse electron subbands are
	\begin{equation}\label{energybands}
	E_{n}\left(k_{x},k_{z}\right)=\sqrt{v^2k_{x}^{2}+v^2 k_{y,n}^{2}+m^{2}\left(k_{z}\right)}
	\end{equation}
	(hole subbands have the opposite sign). 
	In addition to the bulk states, within the interval \mbox{$-k_W<k_z<k_W$}, there are a pair of surface states in the form
	\begin{eqnarray}
	\Psi_{0}\left(y\right)&=&\sqrt{2|m\left(k_{z}\right)|}e^{m\left(k_{z}\right)y/v}\xi_{+}, \nonumber\\
	\Psi_{L}\left(y\right)&=&\sqrt{2|m\left(k_{z}\right)|}e^{-m\left(k_{z}\right)\left(y-L\right)/v}\xi_{-}.
	\end{eqnarray}
	The energy of these states is \mbox{$E_0=v k_x$}, \mbox{$E_L=-v k_x$}, hence, the electrons on the opposite surfaces propagate with opposite group velocity $v_x=\pm v$ in the $x$-direction. Each mode at given $k_z$ contributes to the current in the $x$ direction by the amount
	\begin{equation}
	I^x_0 = ev\quad, \quad\qquad I^x_L = -ev \;
	\end{equation}
	for $-k_W<k_z<k_W$. Integration of $k_z\in[-k_W,k_W]$ produces a total current $\pm evk_W/\pi$ related to the anomalous Hall response in the presence of an electric field in the $y$ direction \cite{Burkov2011}.

	\begin{figure}
		\centering
		\includegraphics[width=0.8\columnwidth]{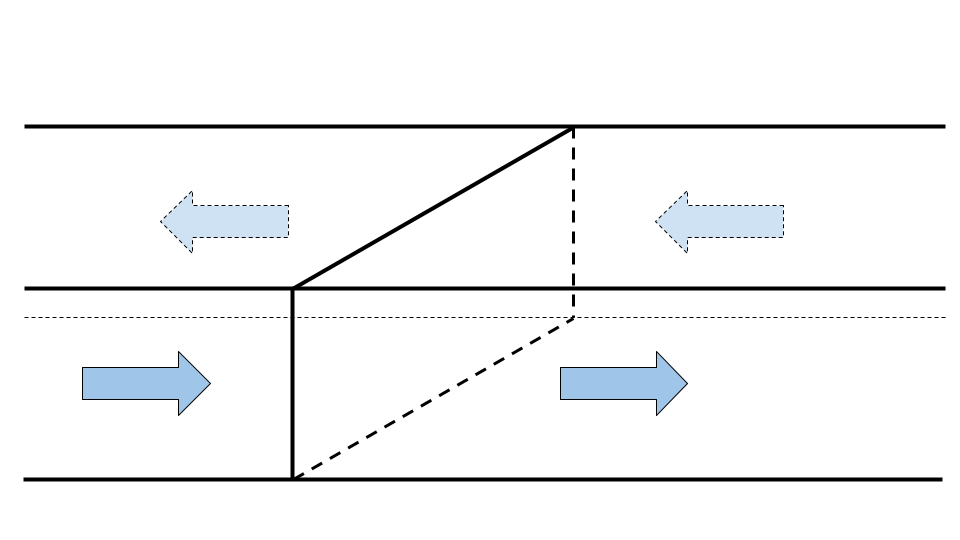}
		\caption{Junction of two magnetic Weyl semimetal slabs with matching Weyl nodes $(\theta=0)$. Perfect transmission of the surface electrons is realized for any value of the potential step.}    \label{fig:Slabjunction0}
	\end{figure}
	
	With a potential step at the interface at $x=0$ between the rotated regions, 
	the momentum $k_x$ is not conserved, but is instead a function of the energy given by
	\begin{eqnarray}
	k_{x,n}&=& \frac{1}{v}\sqrt{\varepsilon_+^2-v^2k_{y,n}^2-m^2\left(k_z\right)} \nonumber\\
	\bar k_{x,n}&=&\frac{1}{v}\sqrt{\varepsilon_-^2-v^2k_{y,n}^2-m^2\left(k_z\right)}
	\end{eqnarray}
	on the two sides. The energy of the surface states is now \mbox{$E_0=v k_x \mp V_0$}, \mbox{$E_L=-v k_x \mp V_0$} 
	on the two sides of the junction, but the group velocity of the electronic states is unaffected.
	In the regime $-V_0<E<V_0$, an electron propagating 
	on the surface at $y=0$ toward the interface can be reflected in any of the bulk 
	modes with amplitude $r_n$ or in the counter-propagating surface state with amplitude $r_L$. 
	Analogously, it can be transmitted in a bulk state with amplitude $t_n$ or in only one of the 
	surface states, the one with matching chirality, with amplitude $t_0$. The state is then written as
	\begin{widetext}
		\begin{equation}\label{slabscatter0}
		\Psi_E=\begin{cases}
		\Psi_{0}\left(y\right)e^{i\varepsilon_+ x/v} + r_L \Psi_{L} \left(y\right) e^{-i \varepsilon_+ x/v} 
		+\sum_{n} r_n \Psi_{n}^{(-)}\left(y\right) e^{-i k_{x,n} x} & x<0, \\
		t_0 \Psi_{0}\left(y\right) e^{i\varepsilon_- x/v} 
		+\sum_{n} t_n \Psi_{n}^{(+)}\left(y\right) e^{i k_{x,n} x} & x>0.
		\end{cases}
		\end{equation}
	\end{widetext}
	In the expression above, when $k_{y,n}$ is sufficiently large, the momenta $k_{x,n}$ and $\bar k_{x,n}$ become imaginary, which accounts for the possibility of having evanescent waves on either side of the interface, with inverse localization length $\kappa_n=-ik_{x,n}$ and  $\bar{\kappa}_{n} =i\bar{k}_{x,n}$.
	Imposing the continuity condition of the wave function at the interface and projecting onto the various bulk states, one readily sees that the only solution is $t_0=1$, with all other coefficients being zero. One then has perfect transmission along the chiral direction, due to the fact that on the other side of the interface there exists a matching state. The two states on the two opposite surfaces behave like "bike lanes", preferential paths for the electron transport across the interface, as illustrated in Fig.~\ref{fig:Slabjunction0}.

	\subsection{Pedestrian crossing}
	An opposite scenario arises when the bulk is depleted of states and the two Fermi arcs on the same side of the slab have opposite chirality: in order for an electron to be either transmitted or reflected (necessarily as a surface state) it must cross the sample in the $y$ direction. To exemplify this situation, we can consider the extreme case in which a sample is contacted with a copy of itself, mirrored in the $yz$ plane, in the absence of a potential bias. We therefore set $\theta=\pi$ and $V_0=0$. 
	Given that for a transparent interface the wave function is continuous at $x=0$, this condition would not be compatible with abruptly reversing the boundary condition for $x<0$ and $x>0$. 
	In order to retain the simplest picture of the Fermi arcs, we require that, sufficiently far from the interface, the states satisfy the boundary conditions \eqref{bc+-}
	for $x<-\ell $, while
	\begin{equation}\label{bc+-j2}
	\begin{cases}
	\sigma^{x}\Psi\left(y=0\right)=-\Psi\left(y=0\right) \;, & \\
	\sigma^{x}\Psi\left(y=L\right)=\Psi\left(y=L\right) \;, & 
	\end{cases}
	\end{equation}
	for $x>\ell$. Here $\ell>0$ is a length scale which models the smooth change in the boundary conditions. Physically, we expect this length to be of the same order of magnitude of the localization length of the interface states. 
	An alternative approach would be to relax the continuity condition at the interface in proximity of the boundaries: as our considerations in this section are mostly qualitative, the two ways of imposing boundary conditions can be considered physically equivalent.
	In fact, if one sample is disconnected from the other, one has a unique Fermi arc which is an eigenstate of $\sigma^x$ on the surface at $y=0$ and an eigenstate of $\sigma^y$ on the surface at $x=0$, provided one is sufficiently far from the origin: around the corner of the sample, the spinor smoothly rotates from one configuration to the other. If two samples with co-propagating Fermi arcs are put in contact, this picture is not altered.

	While the bulk spectrum is unchanged by the $\pi$-rotation, the eigenstates are different on the two sides. In particular, the bulk states are given in \eqref{bulkstatespi}.
	Moreover, a pair of boundary states exists in the region \mbox{$-k_W<k_z<k_W$}, with wavefunction given by
	\begin{eqnarray}\label{piFAs}
	\Psi_{0}\left(x,y\right)&=&\sqrt{2|m\left(k_{z}\right)|/v} \,
	e^{ik_xx+m\left(k_{z}\right)y/v}\xi_{-} , \\
	\Psi_{L}\left(x,y\right)&=&\sqrt{2|m\left(k_{z}\right)|/v} \,
	e^{ik_xx-m\left(k_{z}\right)\left(y-L\right)/v}\xi_{+}, \nonumber
	\end{eqnarray}
	and chiral dispersion, $E= - v k_x$ ($E= + v k_x$) for the state localized at $y=0$ ($y=L$). An electron traveling from $-\infty$ towards the junction cannot be transmitted in the chiral state at $y=0$, as the latter does not support outgoing states. A scattering state can in principle be written in the form \eqref{slabscatter0}, 
	but, following the discussion above, one can only expect such an expression to be valid far from the junction. Clearly, the term $\Psi_0$ has to be substituted by $\Psi_L$ for $x>0$ and the bulk eigenstates by the corresponding eigenstates in the tilted model \eqref{bulkstatespi}.
	One can then divide the system into three regions and impose continuity equations at $x=\pm\ell$ \cite{Cheianov2006}. While the problem is in general complicated in the vicinity of the junction, 
	an especially simple and interesting situation is obtained for $E=0$, for which the only states that propagate in the $x$ direction are the surface states. Given the form of the asymptotic states and the fact that the only bulk states are located at the Weyl nodes, the current which is localized on the surface cannot leak into the bulk on either side. Moreover, the "out" surface states are localized on the far junction: in such a setting, the interface states studied in section \ref{sec:Interfacestates} must provide a sort of "pedestrian crossing" for the electrons, in analogy with what happens in topological insulators \cite{Takahashi2011}. 
	In fact, in the simplified case at hand, the quantization equation \eqref{interfacecontinuity} has the only solution $k_y=0$ and in the region $-k_W<k_z<k_W$, the surface state at $y=0$ transports electric current $I^y=evk_W/\pi$ in the positive $y$ direction, which matches the incoming current from the Fermi arc.
	\begin{figure}
		\centering
		\includegraphics[width=0.8\columnwidth]{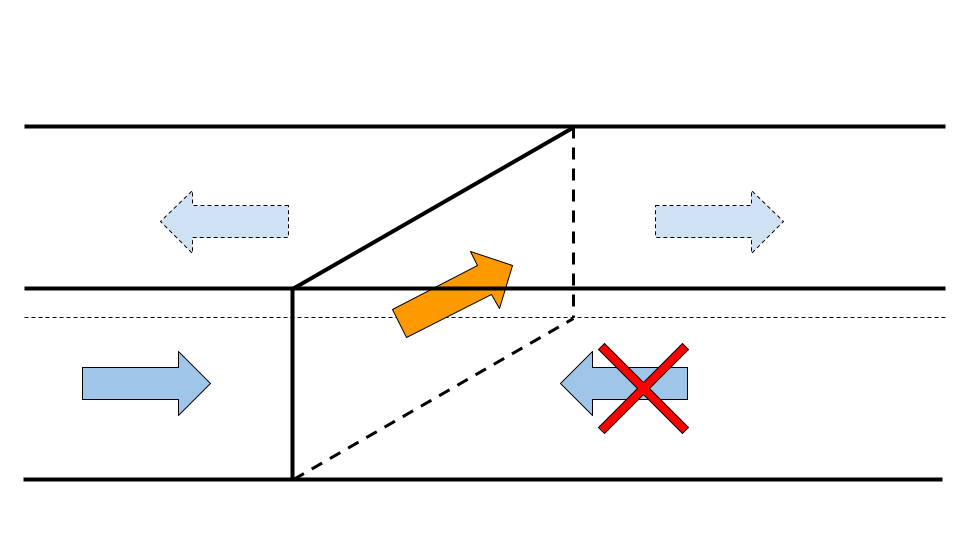}
		\caption{Junction of two magnetic Weyl semimetal slabs, with a $\theta=\pi$ tilt and $V_0=0$. An incoming chiral electron on the surface at $y=0$ must travel along the interface arc.}    \label{fig:Slabjunctionpi}
	\end{figure}
	
	In order to understand what happens at the other end of the interface arc, $y=L$, one can model the interface arc and the Fermi arcs on the two sides as effectively one-dimensional chiral modes entering or exiting a Y-junction. 
	At the junction, we impose the continuity of the boundary condition using \eqref{interfacestates} and \eqref{pisol} for the interface arc and \eqref{piFAs} for the Fermi arcs, reading
	\begin{eqnarray}
	\frac{1}{\sqrt{2}}
	\left(\begin{array}{c}
	-i \\
	1
	\end{array}\right)
	= a_L\xi_- +a_R  \xi_+ \:,
	\end{eqnarray}
	where $a_{L/R}$ are the probability amplitudes for the electron to be transmitted to the left or to the right Fermi arc and satisfy $|a_L|^2+|a_R|^2=1$. One readily obtains the solutions $a_{L/R}=\left(\pm1-i\right)/2$, hence, we conclude that the current splits with equal probability to the two sides of the path, as depicted in Fig.~\ref{fig:Slabjunctionpi}. A current injected from the left side of the interface at $y=0$ can then be measured with half intensity on the right side at $y=L$, which determines a nontrivial signature of the interface states in this specific setup. 
	This splitting of the currents is arguably analogous to the one described in \cite{Abanin2007} for snake states in graphene.
	
	We can now, at least qualitatively, try to understand what happens when the various idealized assumptions are relaxed. 
	While we have studied the simplest situation in order to solve \eqref{interfacecontinuity}, the fact that the Fermi arcs have opposite velocities in the $x$ direction when they arrive at the interface remains true as long as  $\left|\theta-\pi\right|<{\pi}/{2}$, hence we expect this effect to be robust if we vary the tilt angle. On the other hand, for a generic chemical potential, transmission of the Fermi arc excitations into the bulk states on the other side is generically possible, hence, at least part of the current leaks into the bulk.
	Fermi arcs can hybridize with the bulk states, as a consequence of, e.g., scattering by random impurities, although they remain well-defined as long as the bulk Weyl nodes are not gapped out \cite{Armitage2018,Slager2017,Wilson2018}. In fact, the latter are robust against weak disorder \cite{Buchhold2018a,Buchhold2018} and we expect the "pedestrian crossing" scenario to survive in this situation. 
	In real materials, the Fermi arcs do not, in general, describe straight lines in the Brillouin zone \cite{Inoue2016}. Nevertheless, there still is chiral propagation along one direction, hence, the scenario of this section is expected to hold.
	Moreover, multiple pairs of Weyl nodes can be present in the material and, in some instances, their connectivity in the Brillouin zone can depend on the details of the surface \cite{Morali2019}: the bare two-node scenario is clearly not expected to describe this more complicated situation, and it would be interesting to understand whether the simple argument based on the position of the projections of the Weyl nodes will apply.
	Finally, a non-transparent interface, while always allowing for transport from one side of the slab to the other, will reduce the ratio of the current which is found on the other side of the interface against the one on the same side.

	\section{Conclusions}\label{sec:Conclusions}
	We have studied electronic transmission and refraction properties of junctions of
	two magnetic Weyl semimetals in contact via a transparent interface, with mismatched anisotropy axes. We have related the magnitude of the low-temperature conductance and thermopower to the non-trivial topology of the Fermi surface, which features two disconnected or partially overlapping regions. 
	Furthermore, we have studied the momentum refraction at the interface as a function of the incoming momentum and tilt angle, and we have shown that the interface splits the incoming electronic beam according to its chirality. Potentially, this property can find application in the experimental detection of the nature of bulk quasi-particles, i.e., to establish whether there exist chirality-polarized valleys around the Fermi energy, and in the control of the electron trajectories, as this effect is able to produce a beam with a single chirality.
	
	Using a low-energy two-band model, which retains the universal features of the band crossings,  we have established that there exist interface states, connecting the Fermi surfaces around the nodes with the same chirality on the two sides of the slab, whenever the Fermi surfaces do not overlap. We have characterized their chiral transport along the interface and their shape in the Brillouin zone which, although model-dependent, is uniquely fixed by the tilt angle. Arguably, the interface arcs stem from the discontinuity in the local orientation of the Berry curvature at the interface, originated by misaligned Weyl nodes on the two sides. Since the Berry curvature acts on the semiclassical trajectories of the electrons analogously to a magnetic field in momentum space \cite{Son2013}, a discontinuity bears many resemblances with a magnetic field jump in a graphene layer: in particular, the interface states can be seen as three-dimensional analogues of the "snake states" \cite{Ghosh2008}.
	Interestingly, the chirality of the interface arcs implies that they are to some extent robust to backscattering arising from the interaction with, e.g., phonons, while dissipation of the current in the bulk modes is generally possible and most prominent for states in the vicinity of the merging points with the bulk Fermi surface  \cite{Buccheri2021}. It will be interesting to extend the analysis of this paper to related models, including type-II, multi-Weyl and triple-point Weyl semimetals \cite{Fang2012,Soluyanov2015,Zhu2016,Lepori2018}. Possible applications of the work include engineering the interfaces in order to control the path of the interface current \cite{Varnava2021}.

	\begin{acknowledgments}
		We thank C. Mora, M. Burrello and K. Flensberg for interesting discussions and useful feedback. We acknowledge funding  by the Deutsche Forschungsgemeinschaft (DFG, German Research Foundation) under
		Projektnummer 277101999 - TRR 183 (project A02) and Grant No.~EG 96/12-1, 
		as well as within Germany's Excellence Strategy-Cluster of Excellence ``Matter and Light for
		Quantum Computing'' (ML4Q), EXC 2004/1-390534769. 
	\end{acknowledgments}
	
	\appendix
	
	\section{Rotations}\label{sec:rotations}
	The rotation of the Pauli $\sigma$ matrices by an angle $\theta$  around the $x$-axis
	is accomplished by the matrix $ \hat{R}_\theta$, defined by
	\begin{equation}\label{hatR}
	\hat{R}_\theta = e^{-i\frac{\theta}{2}\sigma^x}.
	\end{equation}
	This acts as 
	\mbox{$\sigma_{\theta}^a=\hat{R}_{\theta}\sigma^a \hat{R}_{-\theta}$}, \mbox{$a=x,y,z$}. 
	While $\sigma^x$ clearly commutes with $ \hat{R}_\theta$,  
	the matrices $\sigma^y$ and $\sigma^z$ are brought into the form
	\begin{eqnarray}\label{sigmatheta}
	\sigma_{\theta}^y&=&\cos\theta\sigma^{y} + \sin\theta\sigma^{z},\\
	\sigma_{\theta}^z&=&-\sin\theta\sigma^{y} + \cos\theta\sigma^{z}.
	\end{eqnarray}
	Analogously, while $k_x$ is unaffected by the rotation, 
	the momentum in the $yz$-plane \mbox{$\mathbf{k}=(k_y,k_z)^T$} becomes
	\begin{equation}\label{bfktheta}
	\mathbf{k}_\theta = R_\theta \mathbf{k},
	\end{equation}
	with \mbox{$\mathbf{k}_\theta =(k_{\theta,y},k_{\theta,z})^T$} and
	the rotation matrix \mbox{$R_\theta= \cos\theta \mathbb{I} +i \sigma^y \sin\theta$}.
	Applying these transformations to the bulk Hamiltonian \eqref{HW0}, one readily obtains~ 
	\eqref{HWtheta} for a generic rotation angle $\theta$:
	\begin{equation}
	H_0(k_x,\mathbf{k}) \rightarrow 
	\hat R_\theta H_0(k_x,R_\theta \mathbf{k}) \hat R_{-\theta}= H_\theta(k_x,\mathbf{k}).
	\end{equation}
	The corresponding eigenvectors are
	\begin{eqnarray}\label{bulkeigenvec}
	u_{\theta;k_x,\mathbf{k}} = 
	\sqrt{\frac{1}{2}\left(1-\frac{b_{\theta,z}}{E}\right)}
	\left(\begin{array}{c}
	\frac{vk_{x}-ib_{\theta,y}}{E-b_{\theta,z}} \\
	1
	\end{array}\right)
	\end{eqnarray}
	and have energies $E=E_{\theta,\nu}$ ($\nu=\pm 1$) given in \eqref{Etheta} for the particle/hole branches. 
	(The normalization factor included in Eq.~\eqref{bulkeigenvec} applies only to propagating states.)
	The current density along the $x$-direction carried by the bulk states \eqref{bulkeigenvec} is
	\begin{equation}\label{jxthetabulk}
	j^x_{\theta} 
	=\frac{ev^2k_x}{E_{\theta,\nu}}
	\end{equation}
	and is affected by rotation only through the modified dispersion. (We omit here the volume normalization factor.)

	\section{More on interface states}\label{sec:interapp}
	
	The current density carried by the interface states is evaluated from the wave function \eqref{interfacestates} as
	\begin{equation} \label{jy(x)}
	j^y(x) = ev \Psi^\dagger \sigma^y \Psi =  \frac{2ev^2\mathcal{N}_i^{2}\left(\kappa + k_{y}\right)}{\varepsilon_{+}-m\left(k_{z}\right)}
	e^{-2\tilde{\kappa}\left|x\right|}\;,
	\end{equation}
	where the normalization is
	\begin{equation}\label{internorm}
	\mathcal{N}_i=\sqrt{
		\frac{\kappa \bar \kappa \left(\varepsilon_{+}-m\right)^{2}}
		{2\mathcal{S}\left(\kappa + \bar \kappa \right)
			\left[
			\varepsilon_{+}\left(\varepsilon_{+}-m\right)+v^2\kappa\left(\kappa +k_{y}\right)\right]}}
	\end{equation}
	and $\tilde{\kappa}=\kappa\Theta(-x) + \bar \kappa \Theta(x)$. 
	(Here, $\mathcal{S}$ is a normalization area in the $yz$-plane.)
	The expression \eqref{jy(x)} is exponentially decaying from the junction. 
	Integrating over $x$, one obtains the contribution to the current
	\begin{equation}\label{jy}
	I^y=\frac{ev^2\left(\varepsilon_{+}-m\left(k_z\right)\right)\left(\kappa+k_y\right)}
	{\left[\varepsilon_{+}\left(\varepsilon_{+}-m\left(k_z\right)\right)+v^2\kappa\left(\kappa+k_y\right)\right]\mathcal{S}}
	\end{equation}
	from a mode at given energy and $k_z$. 
	For $\theta\to\pi^\pm$, the wide interface arc between the nodes of negative chirality widens more and more, until its far end is pushed to $k_y\to\mp\infty$. Exactly for $\theta=\pi$, the continuity equation \eqref{interfacecontinuity} reduces to
	\begin{equation}
	\frac{E-m\left(k_{z}\right)}{E+m\left(k_{z}\right)}=
	\frac{k_{y}+\kappa}{k_{y}-{\kappa}}\;,
	\end{equation}
	and admits the two solutions
	\begin{eqnarray}\label{pisol}
	&&  E=vk_y\;,\quad v\kappa=-m\left(k_z\right) \qquad |k_z|<k_W ,\\
	&&  E=-vk_y\;,\quad v\kappa=m\left(k_z\right) \qquad |k_z|>k_W \;. \nonumber
	\end{eqnarray}
	Specializing them at $E=0$ and substituting into \eqref{jy}, we obtain
	\begin{equation}
	I^y=-\,\mbox{sgn}\left(m\left(k_z\right)\right)\frac{ev}{\mathcal{S}}\,.
	\end{equation}
	This expression changes sign at $k_z=\pm k_W$.
	
	\section{Symmetry of interface states dispersion}\label{sec:symmetry}
	
	When $V_0=0$, the equation \eqref{interfacecontinuity} that 
	determines the dispersion relation of the interface states is invariant 
	under a mirror reflection in the plane spanned by $\hat x$ and $R_{-\frac{\theta}{2}}\hat z$. 
	To see this, we rotate the whole system by $\frac{\theta}{2}$ around the $x$ axis, 
	so that the mirror plane coincides with the $xz$-plane. The function $\Phi\left(E,\mathbf{k}\right)$ 
	in Eq.~\eqref{Phifunction} then takes the more symmetric form 
	\begin{equation}
	\Phi \left( E,\mathbf{k} \right) =
	\frac{\left( E - b_{-\frac{\theta}{2},z}(\mathbf{k})\right) 
		\left( -v{\bar \kappa}+b_{\frac{\theta}{2},y}(\mathbf{k})\right)}{
		\left(E - b_{\frac{\theta}{2},z}(\mathbf{k})\right)\left(v\kappa+b_{-\frac{\theta}{2},y}\left(\mathbf{k}\right) \right)}. 
	\end{equation}
	The reflection in the $xz$ plane maps $k_y$ into  $-k_y$. Under this transformation,  we 
	find that
	\begin{equation}
	b_{\pm \frac{\theta}{2},z} \rightarrow  b_{\mp \frac{\theta}{2},z}, \quad 
	b_{\pm \frac{\theta}{2},y} \rightarrow  - b_{\mp \frac{\theta}{2},y}, \quad
	\bar \kappa \leftrightarrow \kappa,
	\end{equation}
	and therefore 
	\begin{equation}
	\Phi \left( E,-k_y,k_z \right) = 1/\Phi \left( E,k_y,k_z \right),
	\end{equation}
	which indeed leaves Eq.~\eqref{interfacecontinuity} invariant. It follows that, in the
	rotated system, $E(\mathbf{k})$ and $v_z(\mathbf{k})$ are even in $k_y$ and $v_y(\mathbf{k})$
	is odd in $k_y$, hence   
	the net current carried by the arc is along the $z$ direction, which is normal to the node displacement 
	vector $\Delta \mathbf{k}_W$.

	\section{Weyl semimetal slab}\label{sec:slabsapp}
	
	In this appendix, we briefly summarize the derivation of the
	eigenstates of $H_0$ and $H_{\theta=\pi}$ in Eqs.~\eqref{HW0} and~\eqref{HWtheta}
	for a slab of finite width $L$ in the $y$ direction and infinite extension in the $x$ and $z$ directions,
	used in Sec.~\ref{sec:Slabs}. Here, the eigenstates are labelled by momentum components
	$k_x$ and $k_z$ and energy $E$. 
	
	We look for eigenstates in the form 
	\begin{equation}\label{fxipm}
	\Psi(y) =   f_{+}\left(y\right)\xi_{+}+f_{-}\left(y\right)\xi_{-},
	\end{equation}
	where
	\begin{equation}\label{xipm}
	\xi_\pm=\frac{1}{\sqrt{2}}\left(\begin{array}{c}
	\pm1  \\
	1
	\end{array}\right)
	\end{equation}
	are the eigenstates of $\sigma^x$. We first consider the case
	$\theta=0$ and apply the Hamiltonian $H_0$
	to the state \eqref{fxipm}. Using
	\begin{equation}
	\sigma^y \xi_\pm=\pm i \xi_\mp \;,\quad 
	\sigma^z \xi_\pm = - \xi_\mp \;,
	\end{equation}
	we obtain the equation
	\begin{eqnarray}
	0&=&
	\Big\{ 
	\left(E-vk_{x}\right)f_{+}+m\left(k_{z}\right)f_{-}+ v f_{-}'\Big\} \xi_{+} \\
	&& +\left\{ \left(E+vk_{x}\right)f_{-}+m\left(k_{z}\right)f_{+}-vf_{+}'\right\} \xi_{-}
	\,.\nonumber
	\end{eqnarray}
	The general solution of the coupled differential equations above reads
	\begin{eqnarray}\label{generalPsi}
	\left(\begin{array}{c}
	f_{-}\\
	f_{+}
	\end{array}\right)&=&
	c_{1}\left(\begin{array}{c}
	vk_{y}\cos\left(k_{y}y\right)-m\left(k_z\right)\sin\left(k_{y}y\right)\\
	\left(E+vk_{x}\right)\sin\left(k_{y}y\right)
	\end{array}\right) \\
	&&+c_{2}\left(\begin{array}{c}
	-\left(E-vk_{x}\right)\sin\left(k_{y}y\right)\\
	vk_{y}\cos\left(k_{y}y\right)+m\left(k_z\right)\sin\left(k_{y}y\right)
	\end{array}\right)
	\nonumber \;,
	\end{eqnarray}
	in which the coefficients $c_1$ and $c_2$ are fixed by the boundary 
	conditions and by the normalization condition, 
	and
	$$
	v^2k_y^2 = E^2-v^2k_x^2-m^2(k_z).
	$$
	Real values of $k_y$ correspond to bulk transverse states, whereas imaginary values $k_y=i\kappa_y$
	correspond to transverse states localized at the surfaces $y=0$ and $y=L$, 
	with inverse decay length $\kappa_y$.
	The boundary conditions \eqref{bc+-} 
	imply \mbox{$f_{-}\left(0\right)=0$} and $f_{+}\left(L\right)\,=\,0$. 
	In turn, these fix $c_{1}=0$ and give the quantization equation
	\begin{equation}\label{bulkquanteq}
	\frac{vk_{y}}{m\left(k_{z}\right)}\cot\left(k_{y}L\right)=-1.
	\end{equation}
	In the large-$|m|L$ limit, the real solutions $k_{y,n}$
	approach the asymptotic values given in  Eq. \eqref{kyn}. 
	The bulk transverse  
	states for a plane wave $e^{\pm i k_xx+ik_zz}$ then read
	\begin{eqnarray}\label{bulkstates0}
	\Psi_{n}^{(\pm)}(y)&=&\mathcal{N}_{n}\Big\{ \left[vk_{y,n}\cos\left(k_{y,n}y\right)+m\left(k_{z}\right)\sin\left(k_{y,n}y\right)\right]\xi_{+}
	\nonumber\\
	&&\qquad\quad -\left(E\mp vk_{x}\right)\sin\left(k_{y,n}y\right)\xi_{-}\Big\}.
	\end{eqnarray}
	For real $k_y$, the normalization factor is, at leading order in $L$,
	\begin{eqnarray}\label{bulknorm}
	\mathcal{N}_{n}^{-1} &=& 
	\sqrt{2LE\left(E\mp vk_x\right)},
	\end{eqnarray}
	and the corresponding subband dispersion relations are
	\begin{equation} \label{bulknrg}
	E_{n}\left(k_{y},k_{z}\right)=\pm\sqrt{v^2k_{x}^{2} + v^2k_{y,n}^{2}+m^{2}\left(k_{z}\right)}.
	\end{equation}
	
	In the same way, we can analyze the case $\theta=\pi$. 
	Using the Hamiltonian \eqref{HWtheta} with $b_\pi^y=vk_y$ and $b_\pi^z=-m\left(k_z\right)$
	and imposing the boundary conditions \eqref{bc+-j2},
	one finds the same quantization condition as in \eqref{bulkquanteq} and the
	bulk transverse states
	\begin{eqnarray}\label{bulkstatespi}
	\Psi_{n}^{(\pm)}(y)&=&\mathcal{N}_{n}\Big\{ 
	\left(E\pm vk_{x}\right)\sin\left(k_{y,n}y\right)\xi_{+} \\
	&&\: +\left[vk_{y,n}\cos\left(k_{y,n}y\right) +m\left(k_{z}\right)\sin\left(k_{y,n}y\right)\right]\xi_{-}\Big\}\;,
	\nonumber
	\end{eqnarray}
	with the spectrum given in \eqref{bulknrg}. 
	The normalization factor $\mathcal{N}_{n}$ is obtained from \eqref{bulknorm} 
	substituting \mbox{$k_x \to -k_x$}. 
	
	\bibliography{WI4}
\end{document}